\documentclass[12pt,a4paper]{article}
\usepackage[german, english]{babel}
\usepackage{a4wide}
\usepackage{amsmath}
\usepackage{amssymb}
\usepackage{ifthen}
\usepackage{epsfig}
\newcounter{fig}   \newcommand{\lbfig}[1]{\refstepcounter{fig}
\label{#1} } 
\newcommand{\vphi}{\varphi}
\newcommand{\Tr}{{\rm Tr}}

\newcommand{\half}{{\textstyle\frac{1}{2}}}

\newcommand{\bea}{\begin{eqnarray}}
\newcommand{\eea}{\end{eqnarray}}
\newcommand{\be}{\begin{equation}}
\newcommand{\ee}{\end{equation}}

\def\diag{\mathop{{\rm diag}}\nolimits}
\def\Laa{\widetilde L}

\def\vecnabla{{\pmb{\nabla}}} 
\newcommand{\da}{\dot\alpha}
\newcommand{\db}{\dot\beta}
\begin{document}

\begin{titlepage}
\vspace*{2cm}

\begin{center}
{\bf\large $SU(N)$ monopoles with and without SUSY}
\vspace{2.0cm}

{\sc\large Ya. Shnir}\\[12pt]
{\it  Institute of Physics, University of Oldenburg}\\
{\it D-26111, Oldenburg, Germany}

\end{center}

\date{~}

\bigskip
\begin{abstract}
These are expanded notes of lectures given at the 
Advanced Summer School on Modern Mathematical Physics (JINR Dubna, July 2005)
and at the  8th International School-Seminar 
``The Actual Problems of Microworld Physics 2005'' (Gomel-Dubna, August 2005). 
I review classical monopole solutions of the $SU(N)$ Yang-Mills-Higgs 
theory. The first part is a pedagogical introduction into 
to the theory of non-Abelian $SU(2)$ monopoles. In the second part 
I discuss a particular case of 
$SU(3)$ theories containing different limits of symmetry breaking. 
It turns out that the multimonopole configurations are natural in a model 
with the gauge group of higher rank. Here I discuss  
fundamental and composite monopoles and consider the limiting situation of the 
massless states. In the last part I briefly discuss 
construction of the $N = 2$ $SU(2)$ supersymmetric monopoles and some of 
the basic properties which are  
connected with the field theoretical aspects of these classical solutions.  
\end{abstract}
\bigskip


\end{titlepage}



\bigskip

\section*{Introduction}
According to some dictionaries,  one meaning of the notion of 
``beauty'' is ``symmetry''. Probably beauty is not entirely in the 
eye of the beholder. It seems to be related to the symmetry of the object. 
From a physical viewpoint this definition is very attractive:  
it allows us to describe a central concept of theoretical physics 
over the last two centuries as being a quest for higher symmetry of Nature. 
The more (super)symmetric the theory, the more beautiful it looks.  
 
Unfortunately our imperfect, (at least at low-energy scale) world is full of nasty 
broken symmetries. This has impelled physicists to try to 
understand how this happens.   
In some cases, it is possible to reveal the mechanism of violation and how the 
symmetry may be recovered; then our picture of Nature becomes a bit more beautiful.

One of the problems of the broken symmetry that we see is that, while there are electric 
charges in our world, their counterparts, {\em magnetic  monopoles}, 
have not been found.      
Thus, in the absence of the monopoles, the  symmetry between electric and 
magnetic quantities is lost. Can this symmetry be regained? If the answer 
will be negative, we have at least to understand 
a reason and it seems plausible that the answer to the question: 
``Why do magnetic monopole not exist?'' is a key to understanding 
the very foundations of Nature. 

In these lectures I focus on the general properties of the 
classical monopole solutions of the $SU(N)$ Yang-Mills-Higgs 
theory.  In the first part I briefly describe properties of $SU(2)$ monopoles, 
both in the BPS limit and beyond. 
In the second part I will review a particular case of 
$SU(3)$ model which allows us to investigate both minimal and 
maximal symmetry breaking. I will then describe some features of N=2 supersymmetric 
monopoles. The last part is 
an introductory account of  $N = 2$ $SU(2)$ supersymmetric monopoles. 
In these lectures I will not consider semiclassical quantization of the solutions. 
In discussion of the low-energy monopole dynamics I will not consider the most powerful  
moduli space approach.   
I also do not venture to discuss the exact duality in supersymmetric 
gauge theories, thus these lectures are restricted to the field-theoretical aspects 
of the monopoles. 
For more rigor and broader discussion I refer the reader to the original publications.

\section{$SU(2)$ Yang-Mills-Higgs model}
\subsection{'t~Hooft - Polyakov monopole}
{\flushright{
{\it "The fox knows many things,\\ 
but the hedgehog knows one big thing"}\\[2pt]
Archiolus\\ 
~
}}

\noindent
We consider non-Abelian classical Lagrangian of a
Yang--Mills--Higgs theory with the gauge group $SU(N)$,
which describes
coupled gauge and Higgs fields:
\be           \label{Lagr}
L = -\frac{1}{2} \Tr F_{\mu\nu}F^{\mu\nu} + \Tr D_\mu \phi D^\mu \phi 
- V(\phi) 
= -\frac{1}{4} F_{\mu \nu}^a F^{a \mu \nu} 
+ \frac{1}{2} (D^{\mu} \phi ^a) (D_{\mu} \phi ^a )  - V(\phi) \ .
\ee
Here, $A_\mu = A_\mu^a T^a$ is an  $SU(N)$ connection with field strenght 
$F_{\mu\nu} = F_{\mu \nu}^a T^a$, $\phi = \phi^a T^a$ and we use
standard normalisation of the Hermitian generators of the gauge group: $\Tr (T^a T^b) =
\frac{1}{2}\delta_{ab}$, $a,b = 1,2,3$, which 
satisfy the Lie algebra 
\begin{equation}
[T^a,T^b]=i\varepsilon_{abc}T^c \ .
\end{equation}
The scalar field $\phi = \phi^a T^a$ transforms in the adjoint representation
of  $SU(N)$ with the covariant derivative defined by 
$D_\mu \phi = \partial _{\mu}\phi + ie [A_\mu,\phi]$. 
The non-zero vacuum expectation value of the scalar field corresponds to the 
symmetry breaking Higgs potential $V(\phi)$ 
\begin{equation}        \label{pot}
V(\phi) = \lambda ( |\phi|^2  - v^2)^2 \, ,
\end{equation} 
where the group norm of the scalar field is defined as 
$ |\phi|^2 = 2 ~{\rm Tr\,} \phi^2 = \phi^a \phi^a$. 

In the simplest non-trivial case of $SU(2)$ group we have $T^a=\sigma^a/2$ and 
$V(\phi) = \lambda (\phi ^a \phi ^a - v^2)^2$. The energy 
of the configuration is minimal if the following conditions are satisfied
\begin{equation}                       \label{higgs-vacuum}
\phi ^a \phi ^a = v^2;\qquad F_{m n}^a = 0;\qquad D_n \phi ^a = 0 \ .
\end{equation}
These conditions define the vacuum.
Note that the very  definition (\ref{higgs-vacuum}) forces the classical vacuum
of the $SU(2)$ Yang-Mills-Higgs theory to be degenerated. 
Indeed, the condition $V(\phi) = 0$
means that $|\phi| = v$, i.e., the set of vacuum values of the Higgs field
forms a sphere $S^2_{\rm vac}$. All the
points on this sphere are equivalent because there is a well defined $SU(2)$
gauge transformation which connects them. If $v^2 \ne 0$ the  $SU(2)$ symmetry 
is spontaneously brocken to $U(1)$.  

Thus, the solutions of the classical field equations map the vacuum manifold ${\cal
M} = S^2_{\rm vac}$ onto the boundary of 3-dimensional space, which is also a
sphere $S^2$.  These maps are charactered by a {\it winding number} $n = 0,\pm
1,\pm 2\dots$ which is the number of times $S^2_{\rm vac}$ is covered by a
single turn around the spatial boundary $S^2$. The celebrated `t Hooft--Polyakov 
solution \cite{Hooft74,Polyakov74} corresponds to the ``hedgehog'' asymptotic 
of the scalar field:
\begin{equation}                                     \label{heudg}
\phi^a ~ \mathrel{\mathop{\longrightarrow}\limits_{r \to \infty}}
\frac{vr^a}{r} \, .
\end{equation}
Such a behavior obviously mixes the spatial and isotopic indices and
defines a single mapping of the vacuum ${\cal M} $ onto the spatial
asymptotic. 

The static regular solution of the corresponding field equations 
was constructed numerically 
by employing of the spherically symmetric Ansatz \cite{Hooft74,Polyakov74}
for the gauge and the Higgs fields, respectively:
\begin{equation}                                 \label{Pola}
\phi ^a = \frac{r^a}{e r^2} H (\xi)\, , \quad A_n^a 
=  \varepsilon _{amn} \frac{r^m }{ er^2} [1 - K (\xi)]\, , \quad A_0^a = 0\, ,
\end{equation}
where $H (\xi)$ and $K (\xi)$ are functions of the dimensionless variable $\xi
= ver$.  

The condition of vanishing covariant derivative of the scalar field on the
spatial asymptotic (\ref{higgs-vacuum}) together with the choice of the
nontrivial hedgehog configuration implies that at $r \to \infty$
\begin{equation}
\partial_n\left(\frac{r^a}{r}\right) 
- e \varepsilon_{abc} A_{n}^b \frac{r^c}{r} = 0.
\end{equation}
The simple transformation
$$
\partial_n\left(\frac{r^a}{r}\right) 
= \frac{r^2 \delta_{an} - r_ar_n}{r^3} 
= \frac{1}{r}\left(\delta_{an}\delta_{ck} 
- \delta_{ak}\delta_{nc}\right) \frac{r_cr_k}{r^2} 
= -\varepsilon_{abc}\varepsilon_{bnk}\frac{r_cr_k}{r^3}
$$
then provides an asymptotic form of the gauge potential 
\begin{equation}                \label{A-as}
A_k^a (r) 
\mathrel{\mathop{\longrightarrow}\limits_{r \to \infty}}
\frac{1}{e} \varepsilon_{ank}\frac{r_n}{r^2}
\end{equation}
This corresponds to the non-Abelian magnetic field
\begin{equation}
B_n^a  
\mathrel{\mathop{\longrightarrow}\limits_{r \to \infty}}
\frac{r_ar_n}{er^4}
\end{equation}
Therefore, the boundary conditions (\ref{heudg}), (\ref{A-as}) are compatible
with the existence of a long-range gauge field associated with an Abelian
subgroup which is unbroken in the vacuum. Since this field decays like
$1/r^2$, which is typical behavior of the Coulomb-like field of a point-like charge, 
and since the electric components of the field strength tensor vanish,
we can recognize a monopole in 
such a ``hedgehog'' configuration with a finite energy. 

\begin{figure}[t]
\begin{center}
\setlength{\unitlength}{1cm}
\begin{picture}(11,8.5)
\put(-0.2,7.8)
{\mbox{\psfig{figure=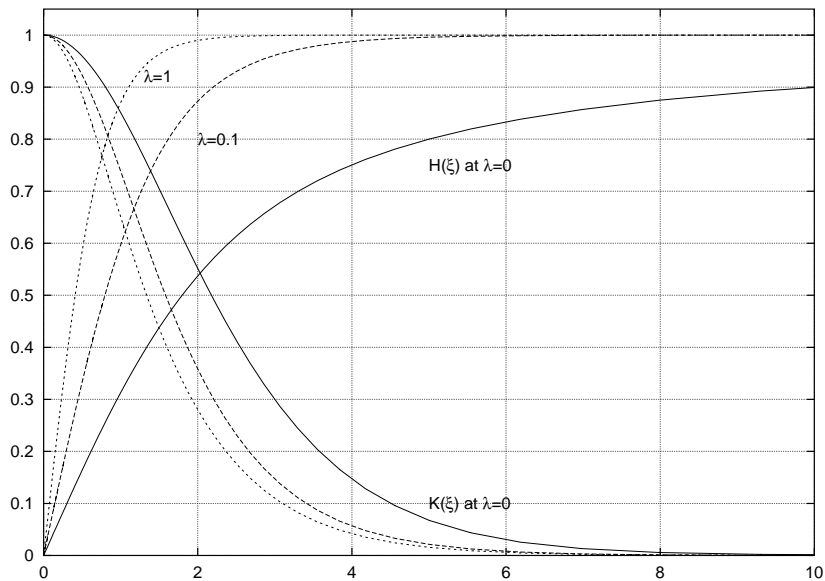,height=11.4cm, angle =-90}}}
\end{picture}
\caption{
The profile functions $K(\xi)$ and $H(\xi)/\xi$ are shown for the 
`t~Hooft--Polyakov monopole at $\lambda =0$, (BPS limit) $\lambda =0.1$ and $\lambda =1$. 
}
\label{fig5.1} 
\end{center}
\end{figure}

The explicit forms of the shape functions of the scalar and gauge
field can be found numerically. It turns out that the functions 
$H(\xi)$ and $K(\xi)$ approach rather fast to the asymptotic values 
(see Fig.~\ref{fig5.1}).

Thus, there is almost vacuum outside of some region of the order of the
characteristic scale $R_c$, which is called the {\it core} of the monopole. 
One could estimate this size by simple arguments
\cite{Preskill84}. The total energy of the monopole configuration consists
of two components: the energy of the Abelian magnetic field outside the core
and the energy of the scalar field inside the core:
$$
E = E_{\rm mag} + E_{s} \sim 4\pi g^2 R_c^{-1} + 4\pi v^2 R_c \sim
\frac{4\pi}{e^2} \left(R_c^{-1} + m_v^2 R_c\right) \, .
$$
This sum is minimal if $R_c \sim m_v^{-1}$. In other words, inside the core at
distances shorter than the wavelength of the vector boson $m_v^{-1} \sim
(ve)^{-1}$, the original $SU(2)$ symmetry is restored. However, outside the
core this symmetry is spontaneously broken down to the Abelian electromagnetic
subgroup.

Furthermore, there is a gauge-invariant definition 
of the electromagnetic field strength tensor ${\cal F_{\mu\nu}}$ \cite{Hooft74}
\begin{equation}
{\cal F}_{\mu\nu} = {\rm Tr} \left\{ \hat \phi F_{\mu\nu}
- \frac{i}{2e} \hat \phi D_\mu \hat \phi D_\nu \hat \phi \right\}
= \hat \phi^a F_{\mu\nu}^a + \frac{1}{e} \epsilon_{abc}
\hat \phi^a D_\mu \hat \phi^b D_\nu \hat \phi^c
\, , \label{Hooft_tensor} 
\end{equation}
where $\hat \phi^a = \phi^a/|\phi|$ is a normalized Higgs field. The 't~Hooft tensor 
corresponds to the magnetic charge of the configuration 
\begin{equation}        \label{g-integral}
g = \frac{1}{v} \int d^2S_n B_n 
= \frac{1}{v}\int  d^2S_n B_n^a \phi^a  
= \frac{1}{v}\int d^3x B_n^a D_n\phi^a \, ,  
\end{equation}
where we make use of the Bianchi identity for the tensor of non-Abelian
magnetic field $D_nB_n^a = 0$.

Note that the solution given by the `t~Hooft--Polyakov ansatz (\ref{Pola})
corresponds to the condition $A_0^a=0$. One could consider a more general case,
where this time component of the vector potential is not equal to zero, but
is also a function of the spatial coordinates \cite{Julia-Zee}:
\begin{equation}                       \label{Julia}
A_0^a = \frac{r^a }{ e r^2} J (r) \, .
\end{equation}
This field configuration corresponds to the non-Abelian {\it dyon}, which has
both magnetic and electric charges. The electric charge of the system of the fields can be
defined as
\begin{equation}        \label{q}
q = \frac{1}{v} \int dS_n E_n 
= \frac{1}{v}\int  dS_n E_n^a \phi^a  
= \frac{1}{v}\int d^3x E_n^a D_n\phi^a \, .  
\end{equation}
Here, we invoked the field equations, according to which
$D_nE_n^a = 0$, and made use of the relation $\varepsilon_{abc}\phi^b
D_0\phi^c = 0$, which is valid for the Ansatz under consideration.
The asymptotic behavior of
the profile function $J(r)$ is very similar to that of the scalar field:
\begin{equation}               \label{bound-dyon}
J(r) \rightarrow 0,\quad  {\rm as}\quad r \rightarrow 0\, , \qquad
J(r) \rightarrow C r \quad {\rm as}\quad r \rightarrow \infty \, .
\end{equation}
The arbitrary constant  $C$ is connected with the electric
charge of the dyon (\ref{q}) \cite{Julia-Zee}. The charge vanishes if $C=0$.

Indeed, substituting the ansatz (\ref{Pola}) into the integral 
(\ref{q}) after some algebra we obtain 
\begin{equation}                          \label{charge-q}
q =  \frac{4 \pi C}{e} = C g \, ,
\end{equation}  
where the magnetic charge of the dyon $g$ is as before given by the formula
(\ref{g-integral}). However, on the classical level there is no reason for the
electric charge (\ref{q}), unlike the magnetic charge, to be quantized and the
constant $C$ in (\ref{charge-q}) remains an arbitrary parameter. 

Finally, we note that the time component of the vector-potential (\ref{Julia})
is in isospace parallel to the direction of the Higgs field. Moreover, one can
consider it as an additional triplet of the scalar fields. This is the 
so-called {\it Julia--Zee correspondence} $\phi^a \rightleftharpoons A_0^a$ .

\subsection{The Bogomol'nyi Limit}

Unfortunately, the system of non-linear coupled differential equations 
on the functions $H(\xi)$ and $K(\xi)$ in general has no analytical solution. 
The only known exception is
the very special case of vanishing scalar potential 
$V=0$  \cite{Bog76,PS75,ColPras77}. This is the
so-called {\it Bogomol'nyi--Prasad--Sommerfield (BPS) limit}.

In the BPS limit of vanishing Higgs potential the 
scalar field also becomes massless and the energy 
of the static field configuration is taking the form
\begin{equation}
E  = \int\left\{ \frac{1}{4} \Tr\left(
\left(\varepsilon_{ijk} F_{ij}\pm D_k \phi \right)^2 \right)
 \mp\frac{1}{2}\varepsilon_{ijk} \Tr\left( F_{ij} D_k \phi \right)	
	\right\} d^3 r \ .
\label{E2}
\end{equation}
Thus, the absolute minimum of the energy corresponds to 
the static configurations which 
satisfy the first order Bogomol'nyi equations:
\begin{equation}
\varepsilon_{ijk} F_{ij} = \pm  D_k \phi 
\label{BPS}
\end{equation}
which are solved by 
\begin{equation}                      \label{BPS-solu}
K = \frac{\xi}{\sinh \xi }; \qquad H = \xi \coth \xi - 1 \ .
\end{equation}  
Definitely,  the solution to the first order BPS equation (\ref{BPS})
automatically satisfies the system of field equations of the second order.

BPS monopoles are very remarkable objects. Let us briefly recapitulate the 
properties of these solutions:
\begin{itemize}
\item[$\bullet$] The BPS equation together with the Bianchi
identity means that $D_nD_n\phi^a = 0$. 
Therefore, the condition
$
D_n\phi^a D_n\phi^a = 
\frac{1}{2}\partial_n\partial_n(\phi^a\phi^a) \, 
$
holds and the energy of the energy of the monopole configuration in the BPS limit is
independent on the properties of the gauge field:
\be 
E = \frac{1}{2} \int d^3 x \partial_n \partial_n (\phi^a \phi^a) = \frac{4\pi v}{e} 
= g v
\ee
If the configuration has both electric and magnetic charges, the monopole mass 
becomes
\be \label{BPS-bound}
M = v\sqrt{g^2 + q^2}
\ee
This yields so-called {\it Bogomol'nyi bound} on the monopole mass. 

\item[$\bullet$] In comparison with the `t~Hooft--Polyakov solution, 
the behavior of the Higgs field of the monopole in the BPS limit has 
changed drastically: as we can see
from (\ref{BPS-solu}), alongside with the exponentially decaying component it
also obtains a long-distance Coulomb tail
\begin{equation}                  \label{as-higgs}
\phi^a \rightarrow v {\hat r}^a - \frac{r^a}{er^2}\quad {\rm as}\quad 
r \rightarrow \infty \, .
\end{equation}
The reason for this is that in the limit $V(\phi)=0$, the scalar field becomes
massless.
\item[$\bullet$]  The long-range monopole-monopole interaction is composed of two
parts originating from the long-range scalar force and the standard
electromagnetic interaction, which could be either attractive or repulsive
\cite{Manton77}. Mutual compensation of both contributions leaves the pair of
BPS monopoles static but the monopole and anti-monopole would interact with
double strength.
\item[$\bullet$] The Bogomol'nyi equation may be treated 
as a three-dimensional reduction of the integrable self-duality equations. 
 Indeed, the Julia--Zee correspondence means that
\begin{eqnarray}                                                   \label{su2}
D_n\phi^a &\rightleftharpoons& D_nA_0^a \equiv F_{0n}^a\, , \nonumber\\
B_n^a = D_n\phi^a &\rightleftharpoons& {\widetilde F}_{0 n} = F_{0n}^a\, .
\end{eqnarray}
Therefore, if we suppose that all the fields are static, the Euclidean
equations of self-duality $F_{\mu\nu}^a = {\widetilde F}_{\mu \nu}^a$ reduce
to the equations (\ref{BPS}) and the monopole solutions in the Bogomol'nyi limit
could be considered as a special class of self-dual fields.
\item[$\bullet$] The analogy between the Euclidean Yang--Mills 
theory and the BPS equations
can be traced up to the solutions. It was shown \cite{Rossi79,Rossi82} that
the solutions of these equations are exactly equal to an infinite chain of
instantons directed along the Euclidean time axis $t$ in $d=4$. More precisely,
the BPS monopole is equivalent to an infinite chain of instantons
having identical orientation in isospace and separated by an interval
$\tau_0 = 2\pi$. An alternative configuration is a chain of correlated
instanton--anti-instanton pairs, which corresponds to an infinite monopole
loop.
\end{itemize}
\subsection{Gauge Zero Mode and the Electric Dyon Charge}
In the BPS limit the Julia--Zee dyonic solutions have a very interesting
interpretation \cite{MantonGibb86,Weinberg79}.
First we note that for the {\it
static} ansatz (\ref{Pola}), (\ref{Julia}) and the choice $A_0 = 0$, the kinetic
energy of the configuration
\begin{equation}                                          \label{kinetic}
T = \int d^3 x \, \Tr\left(E_nE_n + D_0\phi D_0 \phi \right) \, ,
\end{equation}
is equal to zero. Moreover, in this case the Gauss law 
\begin{equation}               \label{Gauss}
D_nE_n -ie \left[\phi, D_0\phi\right] = 0 \, , 
\end{equation}
can be satisfied trivially, with $E_n = D_0\phi =0$.

Let us now consider time-dependent fields $A_n({\bf r},t)$, $\phi^a({\bf
r},t)$, but suppose that their time-dependence arises as a result of a gauge
transformation of the original static configuration:
\begin{equation}         \label{gauge-t-inf}
A_n({\bf r},t) = U({\bf r},t)A_n({\bf r},0)
U^{-1}({\bf r},t) - \frac{i}{e}\, U({\bf r},t)\partial_n U^{-1}({\bf r},t) \, .
\end{equation}
Here, $U({\bf r},t) = e^{ie\omega t}$ with $\omega({\bf r})$ a parameter of the
transformation.  If the time interval $\delta t$ is very small, we can expand
\begin{equation}
U({\bf r},\delta t) \approx 1 + ie\omega \delta t + \dots \, 
\end{equation}
Now it follows from (\ref{gauge-t-inf}) that 
\begin{equation}
A_n({\bf r},\delta t) \approx A_n({\bf r}) 
+ \left(ie[\omega, A_n({\bf r})] - \partial_n \omega \right)\delta t \, ,
\end{equation}
and we have
\begin{equation}           \label{der-gauge}
\partial_0 A_n = ie[\omega, A_n({\bf r})] - \partial_n \omega = -D_n \omega \, .
\end{equation}
In a similar way we obtain for the time-dependence of the scalar field:   
\begin{equation} \label{phi-t}
\partial_0 \phi = ie\left[\omega,\phi\right] \, .
\end{equation}   

These gauge transformations simultaneously affect the time component of the
gauge potential, which for the monopole configuration (\ref{Pola}),
(\ref{Julia}) is a pure gauge:
\begin{equation}
A_0({\bf r},t) = -\frac{i}{e}\,
U({\bf r},t)\partial_0 U^{-1}({\bf r},t) = -\omega \, .
\end{equation}
Since the gauge transformations (\ref{der-gauge}) and (\ref{phi-t}) do not change
the potential energy of the configuration, the parameter $\omega$ can be
identified with the {\it gauge zero mode}. This is one of four collective 
coordinates (they are also called {\it moduli}) of the one-monopole 
configuration \cite{Weinberg79}.  The other three specify the position of the
monopole in space. Their appearance reflects an obvious breaking of
translational invariance of the original Lagrangian (\ref{Lagr}) by the
monopole configuration: the position of the monopole in $\mathbb{R}^3$ can 
be chosen arbitrarily.

However, defined in this way, the gauge zero mode is not physical, since the
gauge transformations (\ref{der-gauge}) and (\ref{phi-t}) do not affect the
non-Abelian electric field:
\begin{eqnarray}
E_n^a &=& \partial_0A_n - D_nA_0 
= -D_n\omega + D_n\omega \equiv 0\, ,\nonumber\\
D_0\phi &=& \partial_0\phi +ie [A_0,\phi] 
= ie\left[\omega,\phi\right] -ie \left[\omega,\phi\right] \equiv 0\, .
\end{eqnarray}
Thus, as before, the Gauss law is satisfied trivially and the kinetic energy
of the monopole (\ref{kinetic}) is still equal to zero.

Now let us suppose that the time-dependence of the fields again appears as a
result of the gauge transformation (\ref{der-gauge}) and (\ref{phi-t}), but that
the corresponding gauge zero mode $(\partial_0A_n, \partial_0\phi)$ 
satisfies the {\it background gauge condition}: 
\begin{equation}                                    \label{back-gauge}
D_n (\partial_0A_n) -ie \left[\phi, (\partial_0\phi)\right] = 0 \, .
\end{equation}
Then the Gauss law (\ref{Gauss}) is satisfied, if $A_0 = 0$ and there is a
non-trivial solution of the equations (\ref{der-gauge}), (\ref{phi-t}) and
(\ref{back-gauge}) \cite{MantonGibb86}, where $\omega$ is proportional to
$\phi$ and an additional time dependence is allowed:
$$
\omega = {\dot \Upsilon}(t)\phi \, ,
$$
which corresponds to the gauge transformation
\begin{equation}              \label{f-t}  
U({\bf r},t) = \exp\{ie\Upsilon(t) \phi({\bf r})\} \approx 1 + ie 
{\dot \Upsilon} \phi \delta t \, .
\end{equation}
Here $\Upsilon(t)$ is an arbitrary function of time. Indeed, in this case we
have $\partial_0A_n = {\dot \Upsilon} D_n\phi $ and $\partial_0\phi = 0$,
and, since in the Bogomol'nyi limit $D_nD_n\phi = 0$, the background gauge
condition (\ref{back-gauge}) is satisfied by the ansatz (\ref{f-t}). However,
this solution corresponds to the generation of a non-Abelian electric field
\begin{equation}
E_n = \partial_0A_n = {\dot \Upsilon}(t)D_n\phi = {\dot \Upsilon}(t)B_n\, ,\qquad
D_0\phi = 0\, ,
\end{equation} 
so the kinetic energy of the monopole (\ref{kinetic}) is no longer zero:
\begin{equation}                           \label{z-mode-kinetic}
\begin{split}
T &= \half{\dot \Upsilon}^2\int d^3x D_n\phi^a D_n\phi^a \\ 
&= \half{\dot \Upsilon}^2\int d^3x B_n^aB_n^a 
= 2 \pi vg {\dot \Upsilon}^2 
= \half M {\dot \Upsilon}^2 \, ,
\end{split}
\end{equation}
where we make use of the definition of the magnetic charge (\ref{g-integral})
and take into account that the mass of the BPS monopole is
$$ M = \frac{4\pi v}{e}\, . 
$$ 

Since the potential energy of the configuration is time-independent, the gauge
transformations (\ref{der-gauge}) and (\ref{phi-t}), supplemented with the
condition $A_0 = 0$, define a physical collective coordinate $\Upsilon(t)$,
that is a gauge zero mode. Its excitation corresponds to the generation of an
electric charge $Q = {\dot \Upsilon} g$. Thus, such a gauge-induced
time-dependence of the fields transforms the monopole into a dyon.

Note that this collective coordinate is an angular variable, which is defined
on a circle $S^1$. Indeed, the points $\Upsilon = 2\pi n $, $n \in \mathbb{Z}$
correspond to the same gauge transformation $ U({\bf r},t) $, which is unity on
the spatial asymptotic \cite{MantonGibb86}.  However, the points $\Upsilon =0$
and, for example $\Upsilon = 2\pi$, correspond to different topological
classes.

To sum up, the one-monopole configuration in the BPS limit could be
characterized by four zero modes (moduli) that form the so-called {\it moduli
space} ${\cal M}_1$.  It is clear from the discussion above that ${\cal M}_1 =
\mathbb{R}^3 \times S^1$. \label{M} \index{Moduli space of monopole}

Note that we can come back to the Julia--Zee description of a dyon
configuration just by inverting the discussion above: we could start from a
system of time-dependent fields and apply the gauge transformations
(\ref{der-gauge}) and (\ref{phi-t}) to compensate for that dependence. The price we
would have to pay, would be the appearance of a non-zero time component of the
gauge potential $A_0$.  This corresponds to the static ansatz (\ref{Julia}).

\subsection{Classical Interaction of Two Widely Separated Dyons}

Now we consider the mechanism of interaction between two widely separated monopoles. 
If they are close enough to each other, the cores overlap and 
we have quite a complicated picture of short-range interactions mediated by the 
gauge and scalar fields.  
However, if we consider well separated monopoles, there is some simplification. We
may suppose that the monopole core has a radius that is much smaller than the
distance between the monopoles.  Moreover, outside of this core the covariant
derivatives of the scalar field vanish and thus the gauge fields obey the free
Yang--Mills equations. This approximation is a standard assumption in the
analysis of monopole interactions.

The result of both analytical \cite{Gold78,Lor79} and variational
\cite{Marg78,Nahm78,Nahm79} calculations confirm a rather surprising
conclusion, first observed by Manton \cite{Manton77}: 
there is no interaction between two BPS monopoles at all, but the
monopole-antimonopole pair attract each other with double strength.

The reason for this unusual behavior is that
the normal magnetostatic repulsion of the two monopoles is balanced by the
long-range scalar interaction: in the BPS limit the quanta of the scalar field
are also massless.

Indeed, we already noted that there is a crucial difference between the
asymptotic behavior of the Higgs field in the non-BPS and the BPS cases: there
is a long-range tail of the BPS monopole
\begin{equation}                                          \label{higgs-as}
\phi^a \rightarrow v {\hat r}^a - \frac{r^a}{er^2}\quad {\rm as}\quad 
r \rightarrow \infty \, .
\end{equation}
The result is that, in a system of two widely separated monopoles, the
asymptotic value of the Higgs field in the region outside the core of the
first monopole is distorted according to Eq.~(\ref{higgs-as}) due to the
long-range scalar field of the other monopole: the mass of the first monopole
will decrease and the size of its core is increased. In other words, the
additional long-range force appears as a result of violation of the original
scale invariance of the model in the BPS limit $\lambda \to 0$. The scalar charge 
of a dyon is simple $Q_D = \sqrt{g^2 + q^2}$ \cite{Manton85} and the corresponding 
Coulomb scalar potential of interaction is simple $\sim - \sqrt{g^2 + q^2}/r$ (recall 
that the scalar interaction is always attractive).

Now we can analyse in more detail the
classical interaction between two dyons that are separated by a large distance
$r$.  Let us suppose that they have identical magnetic charges $g$, but
different electric charges $q_1$ and $q_2$. This problem was studied by Manton
\cite{Manton85} (see also \cite{BakLee94,Bak98}). Again, the situation
is greatly simplified by the assumption of a large separation between them.
Thus, we can neglect the inner structure and consider each dyon as a classical
point-like particle. Since in the BPS limit the dyons possess both electric
and magnetic charges, and the scalar charge $\sqrt{g^2 + q^2}$, as well, the total
interaction of two static dyons is composed of electromagnetic repulsion or
attraction, caused by the electric and magnetic charges, and attraction caused
by the dilaton charges. Thus, the net Coulomb force is
\begin{equation}
{\bf F}_{12} 
= \frac{\bf r}{r^3} \left(g^2 + q_1q_2 - \sqrt{g^2 + q_1^2} \,
\sqrt{g^2 + q_2^2} \right) \ . 
\end{equation}
An additional simplification comes from assuming that the electric charge of
the dyon is much smaller than its magnetic charge. Then an expansion in
$q^2/g^2$ yields
$$
{\bf F}_{12} \approx -\half(q_1 - q_2)^2 \frac{\bf r}{r^3} \, .
$$
In this limit, there is no interaction between two dyons with identical
electric charges. In general, only the relative electric charge of the system
$Q = q_1 - q_2$ enters in the energy of interaction.

Now, let us consider a dyon moving with a velocity ${\bf v}_1$ in the background
field of another dyon, which is placed at rest at the origin \cite{Manton85}.
Since the electromagnetic part of the interaction is described by the Dirac
Abelian potential ${\bf a} = (1-\cos \theta)\ {\bf \vecnabla} \varphi$, 
the canonical momentum of the first dyon is
\begin{equation}
{\bf P} = M {\bf v}_1 + q_1 {\bf A} + g {\bf {\widetilde A}} \, ,
\end{equation}
where the electromagnetic potentials corresponding to the fields of a static
dyon are
$$
{\bf A} = g {\bf a}\, , \qquad
 {\bf {\widetilde A}} = - q_2 {\bf a} \, ,
$$
The scalar electromagnetic potentials that correspond to the static dyon are simply
$$
A_0 = \frac{q_2}{r}\, , \qquad {\widetilde A}_0 = \frac{g}{r} \, .
$$

In addition, we have to take into account one more scalar potential connected with
the scalar charge of the BPS dyon: $\phi = {\sqrt{q_2^2 + g^2}}/r$. As we
have already mentioned above, the effect of this potential is to decrease the mass
of the first dyon as\footnote{Note that a dyon is slightly heavier than a
monopole: $M = M_0 \sqrt{1 + q^2/g^2}$. However, in the case under
consideration, $q \ll g$. Therefore, the difference is of second-order and can
be neglected.}
$$
M \to M - Q_D \phi = M - \frac{1}{r}{\sqrt{q_1^2 + g^2}}{\sqrt{q_2^2 + g^2}} \, .
$$
Collecting all this together, we arrive at the Lagrangian of the motion 
of the dyon in the external field of another static dyon:
\begin{equation}                                        \label{L-two-d}
L_1 = \left(-M + \phi{\sqrt{q_1^2 + g^2}} \right)\sqrt{1 - v_1^2}
+ {\bf v}_1 (q_1 {\bf A} + g  {\bf {\widetilde A}} ) - q_1 A_0 - g {\widetilde A}_0 \, .
\end{equation}

The next step is to incorporate the effect of motion of both dyons. It is 
well-known that, if the background field is generated by a moving source, the
corresponding fields have to be written in the form of Lienard--Wiechert
potentials \cite{BakLee94,Manton85,TTT89}:
\begin{align}                                \label{Lienard}
{\bf A} &= g {\bf a} + q_2 \frac{{\bf v}_2}
{\sqrt{r^2 - [{\bf r} \times {\bf v}_2]^2}} \, , \nonumber\\
 {\bf {\widetilde A}} &= - q_2 {\bf a} 
+ g \frac{{\bf v}_2}{\sqrt{r^2 - [{\bf r} \times {\bf v}_2]^2}}\, , \nonumber\\
A_0 &= \frac{q_2}{\sqrt{r^2 - [{\bf r} \times {\bf v}_2]^2}} 
+ g ({\bf a} \cdot {\bf v}_2)\, , \quad 
{\widetilde A}_0 = \frac{g}{\sqrt{r^2 - [{\bf r} \times {\bf v}_2]^2}} 
-q_2 ({\bf a} \cdot {\bf v}_2) \, , \nonumber\\
\phi &= \frac{\sqrt{q_2^2 + g^2}}
{\sqrt{r^2 - [{\bf r} \times {\bf v}_2]^2}}\sqrt{1 - {\bf v}_2^2} \ .
\end{align}

In the ${\bf a}$-dependent terms we made use of the non-relativistic character
of the motion. Furthermore, in this case we can make the approximation
$\sqrt{r^2 - [{\bf r} \times {\bf v}_2]^2} \approx r$. Substitution of the
potentials (\ref{Lienard}) into the Lagrangian (\ref{L-two-d}) yields, up to
terms of order $q^2 {\bf v}^2$ and ${\bf v}^4$, 
\begin{equation}                               \label{L-two}
L_1 = \frac{1}{2}M {\bf v}_1^2 - \frac{g^2}{2r} ({\bf v}_1 - {\bf v}_2)^2 
+ Qg ({\bf v}_1 - {\bf v}_2) \cdot {\bf a} + \frac {Q^2}{2r} \ .
\end{equation}
It is important that, as we can see from the second term of this expression,
the scalar and magnetic interactions depend on the relative velocity of the
dyons in different ways. The third term here describes the minimal interaction
between the relative charge $Q$ and the magnetic charge $g$, while the last
term is half the standard Coulomb energy of an electric charge $Q$.  (The
other half is associated with the other dyon.)

If we note that all the interaction terms remain the same in the case of the
inverse problem of the dynamics of the second dyon in the background field of
the first one, then the Lagrangian of the relative motion can be obtained
by factorisation of the motion of the center of mass, $M (v_1+v_2)^2/2$, 
from Eq.~(\ref{L-two}) \cite{Manton85}:
\begin{equation}             \label{L-2d}
L = \left(\frac{M}{4}  - \frac{g^2}{2r}\right) {\dot {\bf r}} \cdot 
{\dot {\bf r}} 
+ Qg~{\dot {\bf r}} \cdot  {\bf a} + \frac {Q^2}{2r} \ ,
\end{equation}
where ${\dot {\bf r}} = ({\bf v}_1 - {\bf v}_2)$ is the relative velocity of
the dyons.

Note that the total electric charge is conserved. The corresponding
collective coordinate $q = q_1 + q_2$ can also be factored out with the motion of
the center of mass. 

The equations of motion that follow from the Lagrangian (\ref{L-2d}) are 
\begin{equation}
\begin{split}
           \label{geo-mot}
\left(\frac{M}{2} - \frac{g^2}{r}\right) {\ddot {\bf r}} 
=\frac{g^2}{r^3} \left\{\frac{1}{2} {\bf r} 
( {\dot {\bf r}} \cdot {\dot {\bf r}} ) 
- ({\bf r} \cdot {\dot {\bf r}}) {\dot {\bf r}}\right\} 
+ \frac{Qg}{r^3} [{\dot {\bf r}}
\times {\bf r}] - \frac{Q^2}{2r^3} {\bf r} \, .
\end{split}
\end{equation}
Note that the dynamical equation does not change if we transform the 
Lagrangian of the  relative motion (\ref{L-2d}) as 
\begin{equation}             \label{L-2d-sep}
L = \frac{1}{4}\left(M  - \frac{2g^2}{r}\right) \left({\dot {\bf r}} \cdot 
{\dot {\bf r}} - \frac{Q^2}{g^2}\right) 
+ Qg {\dot {\bf r}} \cdot  {\bf a} \, ,
\end{equation}
where the constant term $MQ^2/4g^2$ is dropped out.

Obviously, for $g=0$, the equation (\ref{geo-mot}) is identical to 
the standard equation of
motion of a charged particle in a Coulomb field. In the general case,
Eq.~(\ref{geo-mot}) can be solved by making use of the corresponding integrals
of motion.  For example, the energy is composed of three terms: normal kinetic
energy, a velocity-depending term originating from the difference between the
dilatonic and magnetic interaction of the dyons, and the standard potential
energy of interaction of the two charges:
\begin{equation}
E = \left(\frac{M}{2}- \frac{g^2}{r}\right) {\dot {\bf r}}^2  - \frac{Q^2}{r} \,  . 
\end{equation}

The second integral of motion is the vector of angular momentum 
\begin{equation}   \label{mom-ch6}
{\bf L} = \left(\frac{1}{2} - \frac{g^2}{Mr}\right){\bf \Laa} 
- Qg \frac{{\bf r}}{r} \, ,
\end{equation}
where ${\bf \Laa} = M [{\bf r} \times {\bf \dot r}]$ is the standard orbital
angular momentum.  Since the relation ${\bf L} \cdot
{\hat {\bf r}} = -Qg = {\it const}.$ holds, the trajectory of 
the relative motion of the dyons
lies on the surface of a cone\footnote{For an analyse of classical 
electromagnetic motion of an electric charge in external field of a static monopole 
see, e.g., \cite{S-book}}. The motion becomes flat only if the magnetic
charge vanishes. 

We have already noted above, that the electric charge of a
static dyon is connected with its fourth cyclic collective coordinate
$\Upsilon $ as $Q \sim {\dot \Upsilon}$.  Excitation of this collective
coordinate can be treated as the appearance of the kinetic energy
(\ref{z-mode-kinetic}).  This analogy now can be generalized in the spirit of
Kaluza--Klein theory \cite{Manton85}. Let us perform the Legendre transform
of the Lagrangian (\ref{L-2d-sep}) 
$$
L({\bf r}, \Upsilon) = L({\bf r}, Q) +  gQ {\dot \Upsilon}\, ,
$$ 
where
\begin{equation}               \label{Q1}
Q \equiv  \frac{2 g^3}{M - \frac{2 g^2}{r}}
\left(\dot \Upsilon + ({\bf a} \cdot {\bf \dot r})\right) \, .
\end{equation}
Then the Lagrangian  (\ref{L-2d-sep}) can be rewritten in the form  
\begin{equation}                 \label{Kal-Kl}
L =  \frac{1}{4}\left(M - \frac{2g^2}{r}\right) 
{\dot {\bf r}} \cdot 
{\dot {\bf r}} + \frac{g^4}{M - \frac{2 g^2}{r}}
\left(\dot \Upsilon + ({\bf a} \cdot {\bf \dot r})\right)^2 \, ,
\end{equation}
which is structure of the Kaluza--Klein Lagrangian 
describing geodesic motion in
the four-dimensional space ${\cal M}_0$ with one compact variable. 
Note that (\ref{Kal-Kl}) does not
depend explicitly on $\Upsilon$. Thus, the corresponding equation of motion is
just the conservation law of the relative electric charge $Q$, Eq.~(\ref{Q1}).
     
To sum up, the relative motion of well-separated BPS dyons is a geodesic
motion in the space  ${\cal M}_0$ governed by the Taub--NUT 
(Newman-Unti--Tamburino) metric
\begin{equation}                             \label{NUT-1}
ds^2 = \left(1 - \frac{2g^2}{Mr}\right) d{\bf r}^2 
+ \frac{\left(\frac{2g^2}{M}\right)^2 }
{ 1 - \frac{2g^2}{Mr}} (d \Upsilon + {\bf a} \cdot d{\bf r})^2 \ . 
\end{equation}
This metric is well-known from general relativity; it was obtained as
early as in 1951.  The Taub--NUT metric corresponds to the
spatially homogeneous solution of the Einstein equations in empty space. 
The length parameter of this metric is $2g^2/M$.
\label{Taub-page}
 
Recall that geodesic motion in a space with the Taub--NUT metric could 
be used only to describe the relative motion of widely separated dyons.  A
general description of the low-energy dynamics of BPS monopoles on the moduli
space is given by the Atiyah--Hitchin metric, whose asymptotic form is the
Taub--NUT metric \cite{AtHit}.

\section{$SU(N)$ monopoles}
So far, we have discussed the monopole field configurations that  
arise as solutions of the classical field equations of the 
simple non-Abelian Yang--Mills--Higgs $SU(2)$
theory. However, it was realized, almost immediately after discovering  the 
't Hooft--Polyakov solution that there are  
other possibilities beyond the simplest non-Abelian model. 
Indeed, the topological
analysis shows \cite{Monastyrski75,Typkin75} that the existence of monopole
solutions is a general property of a gauge theory 
with a semi-simple gauge group $G$, which becomes spontaneously 
broken down  by the Higgs mechanism  
to a residual vacuum subgroup $H$ containing an explicit $U(1)$ factor.
Thus, in general case, the unbroken subgroup $H$ is non-Abelian, that is,  
in addition to the standard electric charge, which is  
associated with the generator of $U(1)$ subgroup, such 
a monopole should also possess some non-Abelian charges.

In the series of papers by E.~Weinberg \cite{Weinberg79,Weinberg80,Weinberg82}
(see also reviews \cite{Weinberg99-1,Weinberg99-2} and recent publications
\cite{Weinberg96,Weinberg96-1,Weinberg96-2,Weinberg98}), many
of the aspects of the monopole solutions in the gauge theory with a 
gauge group of higher rank were discussed. 
Unexpectedly, it turns out that some of these solutions correspond
to massless monopoles. On the other hand, the description of monopoles with
non-Abelian charges provides a new understanding of the duality between electric charges
and monopoles; it becomes transformed into the idea 
of Montonen--Olive duality, which would establish a correspondence between two
different gauge theories, with conventional electrically charged particles  
being treated as physical degrees of freedom within one of these models, 
and monopoles, being considered as fundamental objects within its dual. 

\subsection{Towards Higher Rank Gauge Groups}
Before we start to consider any particular model, 
let us make some general remarks about a generalization of the Georgi--Glashow model. 
We are working with a
Yang--Mills--Higgs theory with an arbitrary simple gauge group $G$ and the
scalar field in the adjoint representation. The corresponding Lagrangian
(\ref{Lagr}) has been written above: 
$$
L = -\frac{1}{2} \Tr\, F_{\mu\nu}F^{\mu\nu} + \Tr\, D_\mu \phi D^\mu \phi -
V(\phi)
$$
Recall that the vacuum manifold $\cal M$ is defined by the relation $|\phi_0|^2 =
v^2$. The stationary subgroup of invariance of the vacuum is $H$ and the topological
classification of the solutions is connected with a map of the space boundary
$S^2$ onto the coset space ${\cal M} =G/H$. In other words, the topological charge of
a magnetic monopole is given by the elements of the homotopy group
$\pi_2(G/H)$.  The problem is to define the stationary subgroup of the vacuum. 

In the simplest case $G =SU(2)$, the residual $U(1)$
symmetry was fixed by the asymptotic of the scalar field; in the unitary gauge we have 
$\phi \to \phi_0 = v T^3 = v Q$. 
This subgroup is identified as an electromagnetic one, i.e., the generator of
the diagonal Cartan subgroup $T^3$ is set to be identical to the operator of the 
electric charge $Q$. Indeed, the potential of
the $SU(2)$ monopole ina singular gauge 
could be constructed by simple embedding of the Dirac
potential into a non-Abelian gauge group: 
$$
{\bf A} = Q \frac{1-\cos\theta}{\sin\theta}{\bf \hat e}_\vphi = Q {\bf A}^{Dirac} \, ,
$$ 
and the color magnetic field of a non-Abelian monopole on the spatial asymptotic 
is given by
$$
B_n = Q \frac{r_n}{r^3} \, .
$$
Therefore, the problem of the construction of monopole solutions for a
higher rank gauge group could be re-formulated as the problem of the  
definition of the matrix $Q$ for a given group \cite{Corr76}.

For $G=SU(N)$, the solution of this problem seems to be rather obvious, because
one can choose the $U(1)$ charge operator from the elements of the corresponding
Cartan subalgebra generated by the  operators $\vec H =
(H_1, H_2 \dots H_{N-1})$ \cite{GoddOlive77}. In other words, the vacuum value
of the scalar field in some fixed direction, for example, in the direction of
the $z$-axis, can be taken to lie in the diagonal Cartan subalgebra of $SU(N)$
\cite{GoddOlive77,Weinberg80}. In this case, 
\begin{equation}                      \label{vacuum-higgs-wein}
\phi_0 = v {\vec h} \cdot {\vec H}  \, , 
\end{equation}
where $\vec h$ is some $N-1$ component vector in the space of Cartan
subalgebra. Thus, the boundary condition on the Higgs field on infinity is
that up to a gauge transformation it is equal to a diagonal matrix of the form
\begin{equation}           \label{diagonal-1}
\phi_0 = {\rm diag}~(v_1, v_2, \dots v_N).
\end{equation} 
Sometimes, this matrix is called a {\it mass matrix},
since the vacuum expectation value for 
the scalar field defines the monopole mass. 
Note that because of the definition of the trace of generators of
the $SU(N)$ group, the sum of all elements $v_i$ is zero.  

Let us consider the pattern of spontaneous symmetry breaking, which is determined 
by the entries of the mass matrix $v_i$. Indeed, the invariant subgroup $H$ 
consists of the transformations that do not change the vacuum $\phi_0$. 
If all the values $v_i$ are different, the gauge symmetry is
{\it maximally broken} and the
residual symmetry group is a maximal torus $U(1)^{N-1}$. In this case,  
it can be thought that in the vacuum we have 
$N-1$ ``electrodynamics'', not just a single one. We can see that 
\begin{equation}
\pi_2\left(\frac{SU(N)}{U(1)^{N-1}}\right) = \pi_1\left(U(1)^{N-1}\right) =
\mathbb{Z}^{N-1} \, ,
\end{equation}
thus, these monopoles are classified by the topological charge  $n = 0,1 \dots N-1$.

Another limiting case is the so-called {\it minimal symmetry breaking}. 
This corresponds to the situation when all but one element of the mass matrix coincide. Then
the group of invariance of the Higgs vacuum is the 
unitary group $U(N-1)$ and one 
can see that there is a single topological charge given by
\begin{equation}
\pi_2\left(\frac{SU(N)}{U(N-1)}\right) = \mathbb{Z} \, .
\end{equation}

An intermediate case of symmetry breaking is that some of the entries of the 
mass matrix are identical. Then the gauge group $G$ is spontaneously
broken to $K \times U(1)^{r}$, where $K$ is a rank $N-1-r$ semi-simple Lie
group. Such a monopole has $r$ topological charges associated with each 
$U(1)$ subgroup, respectively. 

\subsection{Cartan--Weyl Basis and the Simple Roots}

The discussion of the properties of monopoles in a gauge theory of higher rank
is closely connected with notion of the Cartan--Weyl basis
\cite{Lie-text,GoddOlive77}. Let us introduce some notations for the Lie
algebra of an arbitrary\footnote{Recall, that in the case under consideration
$G = SU(N)$, thus the rank of $G$ is $r = N-1$ and $d=N^2-1$, that
is $d-r = N(N-1)$}, simple Lie group of the rank $r > 1$ and the 
dimension $d$.  The Cartan--Weyl basis is constructed by addition  $d-r$ raising
and lowering generators $E_{\vec \beta}$ to the $r$
commuting generators $\vec H$ of the diagonal Cartan subalgebra, 
each for one of the {\it roots} ${\vec \beta}_i = ({\vec \beta}_1, {\vec\beta}_2, \dots \vec\beta_{d-r})$:
\begin{equation}
[H_i,E_{\vec\beta}] = \beta_i E_{\vec\beta};\qquad  [E_{\vec\beta}, 
E_{-{\vec\beta}}] = 2 {\vec \beta} \cdot {\vec H} \, .
\end{equation} 
Here we make use of the internal (vector) product operation in $r$ dimensional
Euclidean {\it root space} $\mathbb{R}^r$.

The advantage of this approach is to put a simple Euclidean geometry into
correspondence to the algebra of Lie group generators.  The roots $\vec
\beta_i$ correspond to the structure constants of a Lie group. These roots,
being considered as vectors in $\mathbb{R}^r$, form a lattice with the following 
properties
\cite{Lie-text,Weinberg82}.
\begin{itemize}
\item[$\bullet$] A semisimple Lie algebra corresponds to every root system.\\ 
\item[$\bullet$] 
The set of roots $\vec \beta_i$ is finite, it spans the entire space $\mathbb{R}^r$ and
does not contain zero elements.\\
\item[$\bullet$] If $\vec\beta$ and $\vec\alpha$ are the roots, the quantity
$2\vec\beta \cdot \vec\alpha /\vec \beta^2 $ is an integer number.\\
\item[$\bullet$] 
If $\vec\alpha$ is a root, the only multiplies of $\vec\alpha$ that are roots
are $\pm \vec\alpha$\\
\item[$\bullet$]
For a root $\vec\beta$ from the set $\vec\beta_i$ and an arbitrary positive
root $\vec\alpha \ne \vec\beta$, the Weyl transformation \index{Weyl reflection} is defined as
\begin{equation}             \label{Weyl-tr}
{\vec \beta}\cdot {\vec\sigma} ({\vec \alpha}) = 
- {\vec\beta} \cdot {\vec \alpha},\quad {\rm where}\quad 
{\vec\sigma} ({\vec \alpha}) = {\vec \alpha} - 2 {\vec \beta} 
\frac{{\vec \beta}\cdot {\vec \alpha}}{{\vec \beta}\cdot {\vec \beta}} \, .
\end{equation}
The set of roots is invariant with respect to 
this transformation. Geometrically, 
the Weyl transformations is a reflection in the hyperplane orthogonal to $\vec
\beta$.\\
\end{itemize}
For non-simple Lie algebra, the roots are split into sets that are
orthogonal to each other.  Thus, it is sufficient to restrict the
consideration to the case of simple Lie groups.

The third property essentially restricts the ambiguities with the choice of the
root vectors. Indeed, if $\vec\alpha$ and $\vec\beta$ are any two roots with
$\vec\alpha^2 \le \vec\beta^2$, then the angle $\gamma$ between these vectors
is no longer arbitrary, because we have
\begin{equation}       \label{sign-root}
\cos \gamma = \pm \frac{n}{2} \frac {| \vec \beta |}{| \vec \alpha |},  \qquad
n \in \mathbb{Z} \, .
\end{equation}
This is possible only if (i) $\vec \beta^2 = \vec \alpha^2,$ ~(ii)~ $\vec
\beta^2 = 2\vec \alpha^2$, and ~ (iii)~ $\vec \beta^2 = 3\vec \alpha^2$. Therefore,
the root diagram for a simple Lie group consists of the vectors of different
lengths with possible values of the angle between these vectors $\pi/6,~ \pi/4,
~ 2\pi/3,~ \pi,~ 3\pi/4$, and   $5\pi/6$.

Note that all these roots can be separated into positive and negative ones,
according to the sign in Eq.~(\ref{sign-root}). One can choose a suitable
basis that spans the root system in such a way that any root $\vec
\beta_i$ can be represented as a linear combination of {\it simple roots} with 
integer coefficients of the same sign, positive or negative. Thus, the
commutative relations of the algebra are determined by the system of the 
corresponding simple roots.

\begin{figure}   
\centering
\includegraphics[height=8.cm]{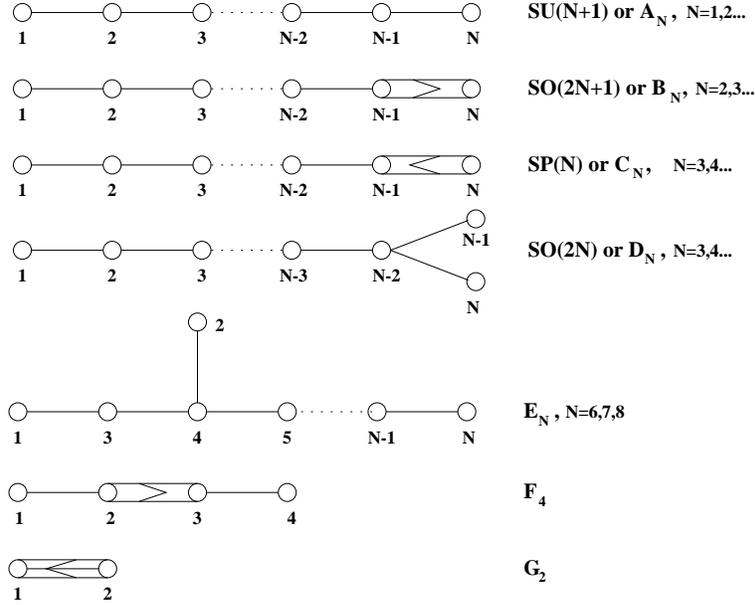}
\caption{
Root diagrams.
\label{fig8.2} }
\end{figure}

The properties of simple roots can be depicted graphically in the form of a
flat graph\footnote{It is hard to resist the temptation to quote V.I.~Arnold
who coined a very nice 
comment concerning the origination of the 
related terminology: 
{\it ``Diagrams of this kind were certainly used 
by Coxeter and Witt, that is
why they are usually called Dynkin diagrams''} \cite{Arn-jk}.}  
as Fig.~\ref{fig8.2}. 
The circles here depict the simple roots.  For any pair of simple roots, we
have $\vec \beta_i \cdot \vec \beta_j \le 0$. Therefore, there are four
possibilities for the angle between the simple roots: $\gamma = \pi/2$ (no
lines on the graph), $2 \pi/3$ (one line), $3\pi/4$ (two lines) and $5\pi/6$
(three lines). The sign `$>$' indicates the length of the simple roots, on one
side of it they are $\sqrt 2$ times longer than on the other.
 
Note that the properties of the simple root basis is related with the symmetry of 
the model. If the group of symmetry $G$ is broken down to the maximal Cartan 
subalgebra, the choice of the simple root basis is unique. However,  
in the case of non-maximal symmetry breaking, an alternative basis may be obtained 
by action of the Weyl reflection. This reflection is actually a 
global gauge transformation from an unbroken non-Abelian subgroup. 
  
A particular choice of the simple roots basis can be specified by means of a
vector $\vec h$ that lies on the root lattice. If this vector 
is not orthogonal to any of the simple roots $\vec \beta_i$, 
the basis is fixed by the condition $\vec h \cdot \vec \beta_i > 0$.

To sum up, the problem of the classification of complex simple (and hence,
semi-simple) algebras is reduced to the problem of the classification of all
non-splittable linearly independent $r$-dimensional systems of root
vectors. This allows us to define a {\it dual} Lie algebra by means of {\it dual}
transformation of the root lattice. 

The dual of a root $\vec \beta$ is defined as
$\vec \beta^* = \vec\beta /\vec \beta^2$ and the duals of the entire set of
simple roots form a dual root lattice of a dual Lie group $G^*$. The dual
lattice is isomorphic to the initial lattice. It is easy to see from the
root diagram (Fig.~\ref{fig8.2}) that, up to rescaling of the root length, the
groups $SU(N)$, $SO(2N)$ and all the exceptional groups are self dual. The
only non-trivial exceptions are the groups $SO(2N + 1)
\rightleftharpoons Sp(N)$, which are dual to each other. 

We illustrate this general description on a particular example of the $SU(3)$
group below.

\subsection{$SU(3)$ Cartan Algebra}

Let us briefly review the basic elements of the $su(3)$ Lie algebra. 
It is given by a set of traceless Hermitian $3 \times 3$ matrices
$$
T^a = \lambda^a/2,\qquad a = 1,2\dots 8 \, , 
$$
where $\lambda^a$ are the standard Gell-Mann matrices. Recall that they are
normalized as $2~\Tr~ T^a T^b = \delta^{ab}$. The structure constants
of the Lie algebra are $f^{abc} = \frac{1}{4} \Tr [\lambda^a, \lambda^b]
\lambda^c$ and in the adjoint representation $(T^a)_{bc} = f^{abc}$.
\begin{figure}[t]   
\centering
\includegraphics[height=7.5cm]{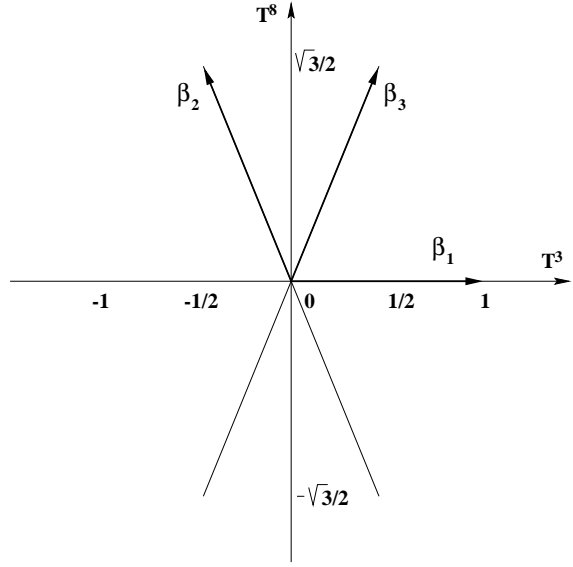}
\caption{
$SU(3)$ simple root self-dual basis.
\label{fig8.3} }
\end{figure}
In the following, we will be especially interested in the diagonal, or Cartan
subalgebra of $SU(3)$.  It is given by two generators
\begin{equation}    \label{H-SU3-ch8}
H_1 \equiv T^3 = \frac{1}{2} \left(\begin{array}{ccc}
1&0&0\\
0&-1&0\\
0&0&0\\
\end{array}\right)\, ,\qquad 
H_2 \equiv T^8 = \frac{1}{2 \sqrt 3}\left(\begin{array}{ccc}
1&0&0\\
0&1&0\\
0&0&-2\\
\end{array}\right) \, , 
\end{equation}
which are composed into the vector ${\vec H} = (H_1, H_2)$. Because the
dimension of the group is $d=8$, the number of positive roots is 3. Taking into
account the restrictions on the angle between the vectors $\vec\beta_i$ and
their length, we can take the basis of simple roots as 
(see~  Fig.~\ref{fig8.3})

\begin{equation}           \label{simple-root}
{\vec \beta}_1 = (1, 0)\, ,\qquad {\vec \beta}_2 = (-1/2, {\sqrt 3}/2) \, .
\end{equation}
The third positive root  is given by the composition of the
first two roots $\vec \beta_3 = \vec \beta_1 + \vec \beta_2 = (1/2,{\sqrt 3}/2)$. 
Since  all these roots have a unit length, our choice corresponds to the 
self-dual basis: ${\vec \beta}_i^* = {\vec \beta}_i$. This allows us to simplify 
the following consideration. 

Note that for any given root ${\vec\beta}_i$ the generators ${\vec \beta}
\cdot {\vec H},~ E_{\pm\beta_i}$ form an $su(2)$ algebra.  The generators
$E_{\pm\beta_i}$ are mentioned above the raising and lowering operators.  Let
us write these generators explicitly in the above-defined basis of the simple roots
(\ref{simple-root}).  
For $\vec \beta_1$, we have 
\bea                   \label{su3-1}
T_{(1)}^3 &=& {\vec \beta}_1 {\vec H} = \frac{1}{2} \left(\begin{array}{ccc}
1&0&0\\
0&-1&0\\
0&0&0\\
\end{array}\right),\\ 
E_{{\vec\beta}_1} &=& \left(\begin{array}{ccc}
0&1&0\\
0&0&0\\
0&0&0\\
\end{array}\right),\qquad 
E_{-{\vec\beta}_1} \equiv \left(\begin{array}{ccc}
0&0&0\\
1&0&0\\
0&0&0\\
\end{array}\right).\nonumber
\eea
For the second simple root ${\vec \beta}_2$, we have
\bea            \label{su3-2}
T_{(2)}^3 &=& {\vec\beta}_2 {\vec H} = \frac{1}{2} \left(\begin{array}{ccc}
0&0&0\\
0&1&0\\
0&0&-1\\
\end{array}\right),\\ 
E_{{\vec\beta}_2}  &=& \left(\begin{array}{ccc}
0&0&0\\
0&0&1\\
0&0&0\\
\end{array}\right),\qquad 
E_{-{\vec\beta}_2}  = \left(\begin{array}{ccc}
0&0&0\\
0&0&0\\
0&1&0\\
\end{array}\right).\nonumber
\eea
The  generators of the $su(2)$ subalgebra that correspond to the  
third  composite root are given by the set of matrices
\bea             \label{su3-3}
T_{(3)}^3 &=& {\vec\beta}_3 {\vec H} = \frac{1}{2} \left(\begin{array}{ccc}
1&0&0\\
0&0&0\\
0&0&-1\\
\end{array}\right),\\ 
E_{{\vec\beta}_3}  &=& \left(\begin{array}{ccc}
0&0&1\\
0&0&0\\
0&0&0\\
\end{array}\right),\qquad 
E_{-{\vec\beta}_3}  = \left(\begin{array}{ccc}
0&0&0\\
0&0&0\\
1&0&0\\
\end{array}\right).\nonumber
\eea
Clearly, the set of matrices $T^a_{(k)}$,  $k=1,2,3$, which includes $ T_{(k)}^3$ of 
Eqs.~(\ref{su3-1}), (\ref{su3-2}) and (\ref{su3-3}), and  
\begin{equation}
T^1_{(k)} = \frac{1}{2}\left(E_{{\vec\beta}_k} + E_{-{\vec\beta}_k}\right), \qquad T^2_{(k)} = 
\frac{1}{2i}\left( E_{{\vec\beta}_k} -  E_{-{\vec\beta}_k}\right)
\end{equation}
satisfy the commutation relations of the $su(2)$ algebras associated with the
simple roots ${\vec \beta}_1, ~{\vec\beta}_2$ and  ${\vec \beta}_3$, respectively.

Let us consider the $su(2)$ subalgebra associated with the first simple root.
If we supplement it by the $U(1)$ hypercharge operator, 
which is connected with the element of the Cartan subalgebra as
$$
Y = \frac{2}{\sqrt 3} T^8 =  \frac{1}{3} \left(\begin{array}{ccc}
1&0&0\\
0&1&0\\
0&0&-2\\
\end{array}\right),
$$
we arrive to the $u(2)$ algebra generated by operators $T_{(1)}^a, Y$. By analogy
with the Euler parameterization of the $SU(2)$ group, an element of corresponding
$U(2)$ transformation can be written as
\begin{eqnarray}     \label{R}
&R_{{\beta}_1}(\gamma,\vphi,\theta,\psi) = R_Y(\gamma) R_3(\vphi)R_2(\theta)
R_3(\psi) = e^{i \gamma Y} e^{i \vphi  T_{(1)}^3 }e^{i \theta T_{(1)}^2} e^{i \psi T_{(1)}^3}
\nonumber\\[3pt]
&=\left(\begin{array}{ccc} e^{i\frac{\gamma}{3}}&0&0\\
0&e^{i\frac{\gamma}{3}}&0\\ 0&0&e^{-\frac{2i\gamma}{3}}\\
\end{array}\right) \left(\begin{array}{ccc}
\cos \frac{\theta}{2} e^{\frac{i}{2}(\vphi+\psi)}~&~\sin \frac{\theta}{2}
e^{\frac{i}{2}(\vphi-\psi)} &~0\\
-\sin \frac{\theta}{2}
e^{\frac{i}{2}(\psi-\vphi)}~&
~\cos \frac{\theta}{2} e^{-\frac{i}{2}(\vphi+\psi)} &~0\\
0&0&~1\\
\end{array}\right)\, , \nonumber\\ 
\end{eqnarray}
where the angular variables are changing within the intervals $0 \le \gamma <
2\pi$, $0 \le \vphi < 2\pi$, $0 \le \theta < \pi$, and $0 \le \psi <
4\pi$. Here the points corresponding to the values $\gamma$ and $\gamma +
\pi$;~$\psi$ and $\psi + 2 \pi$ are pairwise identified, which corresponds to
the $\mathbb{Z}_2$ subgroup.

An alternative choice is  
\begin{eqnarray}     \label{R-2}
&R_{{\beta}_2}(\gamma,\vphi,\theta,\psi) =
e^{i \gamma Y} e^{i \vphi  T_{(2)}^3}e^{i \theta T_{(2)}^2} e^{i \psi T_{(2)}^3}\\[3pt]
&=\left(\begin{array}{ccc}
e^{i\frac{\gamma}{3}}&0&0\\
0&e^{i\frac{\gamma}{3}}&0\\
0&0&e^{-\frac{2i\gamma}{3}}\\
\end{array}\right) \left(\begin{array}{ccc}
1~&0&0\\
0~&\cos \frac{\theta}{2} e^{\frac{i}{2}(\vphi+\psi)}~&~\sin \frac{\theta}{2}
e^{\frac{i}{2}(\vphi-\psi)}\\
0~&-\sin \frac{\theta}{2}
e^{\frac{i}{2}(\psi-\vphi)}&
~\cos \frac{\theta}{2} e^{-\frac{i}{2}(\vphi+\psi)}\\
\end{array}\right) \nonumber \, .
\end{eqnarray}

In other words, the basis of the simple roots $\vec \beta_1$, $\vec \beta_2$ corresponds
to two different ways to embed the $SU(2)$ subgroup into $SU(3)$. The upper
left and lower right $2 \times 2$ blocks correspond to the subgroups
generated by the simple roots $\beta_1$ and $\beta_2$, respectively. The
third composite root ${\vec \beta}_3$ generates the $SU(2)$ 
subgroup, which lies in the corner elements of the $3\times 3$ matrices 
of $SU(3)$.

\subsection{Construction of the {\bf$SU(3)$} Monopoles}
We consider the Yang--Mills--Higgs system, that is governed
by the Lagrangian (\ref{Lagr}) with a gauge group $SU(3)$, as an
explicit example of construction of the monopole solutions in a model with
a large symmetry group.  Thus, the Higgs field $\phi = \phi^a T^a$ is taken 
in the adjoint representation of $SU(3)$, which is given by the set of 
Hermitian matrices $T^a$. 

Unlike the original $SU(2)$ `t~Hooft--Polyakov monopole solution, 
the vacuum manifold ${\cal M}$ of the $SU(3)$ Yang--Mills--Higgs theory is a 
sphere $S^7_{vac}$ in eight-dimensional space. Thus, the topological
classification of the solutions is related with the mapping of the spatial asymptotic
$S^2$ onto coset space ${\cal M} = SU(3)/H$, where $H$ is a residual symmetry
of the vacuum. Another, not so obvious, difference is that now all the points
of the vacuum manifold ${\cal M}$ are not identical up to a gauge
transformation, because the action of the gauge group $SU(3)$ is not transitive.

Thus, in order to classify the solutions, we have to define the unbroken
subgroup $H$.  According the general relation (\ref{vacuum-higgs-wein}), the
asymptotic value of the scalar field in some fixed direction can be chosen to
lie in the Cartan subalgebra, i.e.,
\begin{equation}                      
\phi_0 = v {\vec h} \cdot {\vec H} \, . 
\end{equation}
Clearly, this is a generalization of the $SU(2)$ boundary condition 
$\phi_0 =  v \sigma_3/2$. 
To fix the basis of simple roots, we suppose that all these roots have a
positive inner product with $\vec h$.

Furthermore, if the monopole solution obeys the Bogomol'nyi equations, in the 
direction chosen to define $\phi_0$, 
the asymptotic magnetic field of a BPS monopole is also of the form 
\begin{equation}                  \label{non-bgml-charg}
B_n = {\vec g} \cdot {\vec H} \frac{r_n}{r^3} \, .
\end{equation}
Here the magnetic charge $ g = \vec g \cdot {\vec H} $ is defined as a vector
in the root space \cite{Eng76,GoddOlive77}. 

The principal difference
from the $SU(2)$ model is that the magnetic charge and the topological charge are now
no longer identical. 
Indeed, it is seen that the $SU(3)$ magnetic charge is labeled by two integers. The
charge quantization condition may be obtained from the requirement of
topological stability.  In other 
words, the phase factor must be restricted by  \cite{Eng76,GoddOlive77}
\begin{equation}  \label{top-quantization}
\exp\{i e {\vec g} \cdot {\vec H}\} = 1 \, .
\end{equation}
General solution of this equation is given by a condition that the 
vector charge ${\vec g}$ lies on the dual root lattice 
\cite{Corr76-2,Eng76,GoddOlive77}:
\begin{equation}             \label{wein-quant}
{\vec g} =\frac{4 \pi}{e}\sum\limits_{i=1}^{r}n_i \vec\beta_i^*  = 
\frac{4 \pi }{e} 
\left(n_1 {\vec \beta_1}^* + n_2 {\vec \beta_2}^*\right) = g_1 
{\vec \beta_1}^* + g_2 {\vec \beta_2}^* \, ,
\end{equation}
where $n_1$ and $n_2$ are non-negative integers, and 
${\vec g}_1, {\vec g}_2$ are the magnetic 
charges associated with the corresponding simple roots.

Recall that a special feature of the basis of simple roots (\ref{simple-root}) 
is that it is self-dual: ${\vec \beta_1}^* = {\vec \beta_1};~ {\vec \beta_2}^* = {\vec
\beta_2}$.  Thus, in terms of the explicitly defined roots (\ref{simple-root}),  
we have   
\begin{equation}        \label{su3quant}
g = {\vec g} \cdot {\vec H} = 
\frac{4 \pi }{e} \left[ \left(n_1 - \frac{n_2}{2}\right) H_1
+ \frac{\sqrt 3}{2} n_2 H_2\right] \, .
\end{equation}

These relations show that a magnetic charge is not so trivially quantized  
as in the $SU(2)$ model; the latter has a little bit too much symmetry. 
Thus, the question is, if both of the numbers $n_1,n_2$ can be set 
into correspondence with some topological charges. Evidently, the answer 
depends on the pattern of the symmetry breaking.

\subsubsection{$SU(3)\to U(1)\times U(1)$: Maximal Symmetry Breaking} 
Let us consider two situations that are possible for the $G=SU(3)$ \cite{Weinberg80}. 
\begin{figure}[t]   
\centering
\includegraphics[height=5.5cm]{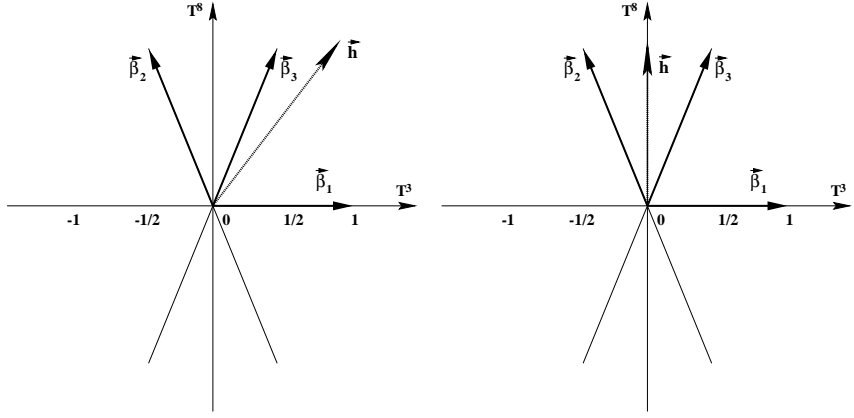}
\caption{
Orientation of the Higgs field in the $SU(3)$ root space, 
which corresponds to the pattern of the maximal symmetry breaking 
$SU(3)\to U(1)\times U(1)$ (left) and minimal symmetry breaking 
$SU(3)\to U(2)$ (right).
\label{fig8.4} }
\end{figure}
%
If the Higgs vector ${\vec h}$ is not orthogonal to any of the simple roots
$\vec \beta_i$ (\ref{simple-root}), there is a unique set of simple roots with 
positive inner product with ${\vec h}$. Thus, the symmetry is maximally broken to the
maximal Abelian torus $U(1)\times U(1)$ 
(see the root diagram of Fig.~\ref{fig8.4}, left). 

If the inner product of ${\vec h}$ and either of
the simple roots is vanishing (see Fig. \ref{fig8.4}, right,
where ${\vec h} \cdot {\vec \beta}_1 = 0$), there are two choices of the basis
of simple roots with positive inner product with  ${\vec h}$, which are related
by Weyl reflections. We shall discuss this type of 
minimal symmetry breaking below.

In the case of maximal symmetry breaking, the topological consideration shows
that
\be        \label{max-hom}
\pi_2\left(\frac{SU(3)}{U(1)\times U(1)}\right) = \mathbb{Z}_2 \, .
\ee
Thus, both of the numbers $n_1,n_2$ have the meaning of topological
charges. Indeed, we define a magnetic charge as a winding number given by the
mapping from a loop in an arbitrary Lie group into the circle on the spatial
asymptotic \cite{Lubkin63}. Then the topological charge of a non-Abelian
monopole is given by the integral over the surface of sphere $S^2$ (cf. the
definition (\ref{g-integral}))
\begin{equation}                     \label{g-int-su}
G = \frac{1}{v} \int dS_n \Tr (B_n \phi_0) 
=  {\vec g} \cdot {\vec h} \, .
\end{equation}
Therefore, if ${\vec h}$ is orthogonal to the root $\vec \beta_1$, only one
component of ${\vec g}$ may be associated with the topological charge. Otherwise, 
there are two topological integers that are associated with a monopole.

The definition of topological charge (\ref{g-int-su}) can be used 
to generalize the Bogomol'nyi bound (\ref{BPS-bound}) for the  
$SU(N)$ monopoles. If we do not consider degrees of freedom that are 
related with  electric charges of the configuration, it becomes 
\begin{equation}         \label{BPS-non-abel}
M = v | G | =   
\frac{4 \pi v }{e}\sum\limits_{i=1}^{r}n_i  \left(\vec h \cdot \vec\beta_i^*\right) =
\sum\limits_{i=1}^{r}n_i M_i \, ,
\end{equation}
where $M_i = \frac{4 \pi v }{e}~\vec h \cdot \vec\beta_i^*$ and we suppose that the 
orientation of
the Higgs field uniquely determines a set of simple roots that satisfies the
condition $\vec h \cdot \vec\beta_i^* \ge 0$ for all $i$. Thus, it looks like 
there are $r$ individual monopoles of masses $M_i$. 

Moreover, there is an obvious analogy between the relation (\ref{wein-quant})
and the definition of a 
magnetic charge of a multimonopole configuration of the $SU(2)$ model, which is 
given by the sum over separate monopoles with a minimal charge. 
Thus, the question arises, if the
monopole solutions of a higher rank gauge theory may also  be understood as a
composite system of a few single monopoles with a minimal charge, masses $M_i$
and characteristic sizes of cores
\begin{equation}
R_c^i \sim (ve {\vec h} \cdot {\vec \beta}_i)^{-1} 
\end{equation}
correspondingly. A very strong argument in support of this conclusion is
given by a direct calculation of the number of zero modes on the monopole background
\cite{Weinberg79,Weinberg80,Weinberg82}.

To analyse the situation better, let us return to 
a system of spherically symmetric $SU(3)$ monopoles 
in the basis of simple roots  (\ref{simple-root}).
Such a configuration  
can be constructed by a simple embedding \cite{Bais78-2,Corr76,Kunz93}. 
The recipe is obvious: we have to choose one of the simple
roots having a positive inner product with the scalar field, e.g.,
$\vec \beta_1$, and embed the `t Hooft--Polyakov solution into the 
corresponding $SU(2)$ subgroup. For example, embedding into the left upper corner  
$SU(2)$ subgroup  
defines the $\beta_1$-monopole that is characterised 
by the vector charge $\vec g =
(1,0)$ and the mass $M_1$, while the embedding into the lower right corner  
$SU(2)$ subgroup  
defines the $\beta_2$-monopole with  the vector charge $\vec g =
(0,1)$ and the mass $M_2$.

Embedding of the spherically symmetric $SU(2)$ monopole 
along composite root $\beta_3$ gives a $\vec g = (1,1)$ monopole with the magnetic 
charge 
$$
g = \vec g \cdot  \vec H = \frac{1}{2} H_1 + \frac{\sqrt 3}{2}H_2 \, .
$$ 
Moreover, its mass is equal to the sum of masses of 
the  $\beta_1$-monopole and $\beta_2$-monopole:
$M_1+M_2$. 

The analysis based on the index theorem shows
\cite{Weinberg80} that this configuration is a simple
superposition of two other fundamental solutions and can be continuously
deformed into a solution that describes two well-separated single $\beta_1$ and $\beta_2$ 
monopoles. We shall check this conclusion by making use of another arguments below.
Note that if the Higgs field is oriented
along the composite root, i.e., if $\vec h = \vec \beta_3$, two fundamental BPS
monopoles have the same mass:
$$
M_1 = M_2 = \frac{2 \pi}{e} \, ,
$$
which is half of the mass of the $\beta_3$ monopole. 
In all other cases, this degeneration is lifted and one of the monopoles is
heavier than the other one.

\subsubsection{Spherically Symmetric $SU(3)$ Non-BPS Monopoles}

To construct the 
embedded monopoles, we must  take into account that the generators  $T^a_{(i)}$ 
of an $SU(2)$ subgroup commute with the invariant component of the Higgs field
$$
\phi^{(h)} =  \left({\vec h} - \frac{\vec h \cdot \vec\beta_i}{\beta^2} 
\vec \beta_i \right){\vec H}, \qquad [T^a_{(i)}, \phi^{(h)}] = 0 \, .
$$

Thus, an embedded $SU(2)$ monopole is defined as \cite{Bais78-2}:
\begin{equation}          \label{emb-Bais}
A_n = A_n^a {T}^a_{(i)}, \qquad \phi = \phi^a {T}^a _{(i)} + v \phi^{(h)} \, . 
\end{equation}
The additional invariant term $\phi^{(h)}$ is added to the Higgs field to 
satisfy the boundary conditions on the spatial asymptotic. 
In our basis of the simple roots, we can write  
\begin{equation}  \label{h-inv}
\begin{split}
\vec \beta_1:~~~& \phi^{(h)} = \frac{h_2}{2\sqrt 3}  \left(\begin{array}{ccc}
1&0&0\\
0&1&0\\
0&0&-2\\
\end{array}\right),\\
\vec \beta_2:~~~& \phi^{(h)} = \frac{1}{4}\left(h_1 + \frac{h_2}{\sqrt 3}\right)  \left(\begin{array}{ccc}
2&0&0\\
0&-1&0\\
0&0&-1\\
\end{array}\right),\\
\vec \beta_3:~~~& \phi^{(h)} = \frac{1}{4}\left(h_1 - \frac{h_2}{\sqrt 3}\right)   \left(\begin{array}{ccc}
1&0&0\\
0&-2&0\\
0&0&1\\
\end{array}\right) \, .
\end{split}
\end{equation}
Clearly, the embedding (\ref{emb-Bais}) is very convenient for obtaining 
spherically symmetric 
monopoles \cite{Weinberg82}. It is also helpful for examing the 
fields and low-energy dynamics of 
the charge two BPS monopoles \cite{Irwin97}.  Depending on the boundary conditions and 
pattern of the symmetry, some other ans\"atze 
can be implemented to investigate static monopole 
solutions, such as, for example, the harmonic map ansatz \cite{Sutcliffe99} 
that was used to construct
non-Bogomol'nyi $SU(N)$ BPS monopoles.

In our consideration \cite{S-book,S03}, which  is not restricted to the case of BPS limit, 
we shall consider  
ans\"atze for the Higgs field of a spherically symmetric $\beta_i$ monopole configuration. 
Depending
on the way of the $SU(2)$-embedding, it can be taken\footnote{The first of these Ans\"atze 
(in a different  basis of the simple roots) was already used in \cite{Brihaye01,Kunz93}.} 
as a generalization of the the embedding (\ref{emb-Bais}) 
\begin{equation}        \label{h-1}
\begin{split}
\vec \beta_i:~~~&\phi(r) = \Phi_1(r) \tau^{(i)}_r + \frac{\sqrt 3}{2} \Phi_2 (r) D^{(i)},\\
A_r =& 0;\quad A_\theta = [1-K(r)] \tau^{(i)}_\varphi,\quad
A_\vphi = -\sin\theta [1-K(r)] \tau^{(i)}_\theta \, ,\\
\end{split}
\end{equation}
where $i=1,2,3$, and we make use of the $su(2)$ matrices 
$\tau^{(i)}_r = \left(T_{(i)}^a {\hat r}^a\right)$, 
$\tau^{(i)}_\theta = \left(T_{(i)}^a {\hat \theta}^a\right)$ and 
$\tau^{(i)}_\varphi = \left(T_{(i)}^a {\hat \varphi}^a\right)$. The 
diagonal matrices $D^{(i)}$, which define the embedding along 
the corresponding simple root, are just the $SU(3)$ hypercharge  
\be 
D^{(1)} \equiv Y = \frac{2}{\sqrt 3}H_2 = \frac{1}{3}\diag(1,1,-2),
\ee
the $SU(3)$ electric charge operator 
\be \label{Q-su}
D^{(2)} \equiv Q = T^3 + \frac{Y}{2} = H_1 + \frac{H_2}{\sqrt 3} = \frac{1}{3}\diag(2,-1,-1), 
\ee
and its conjugated operator 
$$
D^{(3)} \equiv {\tilde Q} = T^3 - \frac{Y}{2} = \frac{1}{3}\diag(1,-2,1)\, .
$$

\begin{figure}[t]   
\centering
\includegraphics[height=5.5cm]{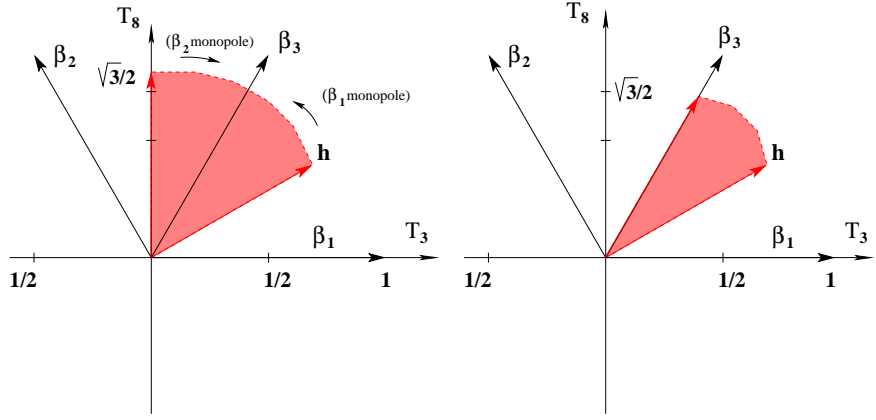}
\caption{
Domains of rotations of the vector 
$
\vec h $ 
for fundamental and composite $SU(3)$ monopoles.}
\label{fig8.5} 
\end{figure}

Inserting the ansatz (\ref{h-1}) into the rescaled Lagrangian (\ref{Lagr}), we can obtain the 
set of the variational equations which may be solved numerically for the range of 
values of vacuum expectation values  $(\Phi_2)_{vac}$.  According to the boundary conditions
on the spacial asymptotic, which fixes the inner product of the vector $\vec h$ 
with all roots has to be non-negative 
for any embedding, the  increasing constant  $(\Phi_2)_{vac}$ 
results in 
rotation of the vector $\vec h$ in the root space, as shown in Fig.~\ref{fig8.5}. 
However, for a single fundamental 
$\beta_i$ monopole,   
$(\vec h\cdot \vec \beta_i) \ge 0$ if $(\Phi_2)_{vac} \ge {1}/{2}$, and in this case 
$(\Phi_2)_{vac}$ have to be restricted as $(\Phi_2)_{vac}\in [1/2;1]$, whereas for a configuration 
embedded along the composite root $\vec \beta_3$, we have $(\Phi_2)_{vac} \le {1}/{2}$ \cite{S03}.

The physical meaning of the 
third of the ans\"atze for the scalar field (\ref{h-1}) becomes clearer, if we note that 
on the spatial asymptotic this configuration really corresponds to the   
Higgs field of two distinct fundamental monopoles, 
$(1,0)$ and $(0,1)$, respectively. Indeed, outside of cores of these monopoles in the Abelian gauge, the 
scalar field can be written as the superposition:
$$
\phi (r\to \infty) =  v_1 T_{(1)}^3 + v_2 T_{(2)}^3 =  \frac{1}{2}\left(\begin{array}{ccc}
v_1&0&0\\
0&v_2-v_1&0\\
0&0&-v_2\\
\end{array}\right) \, ,
$$
where the Higgs field of the $\beta_1$ and $\beta_2$ monopoles takes the  
vacuum expectation values $v_1, v_2$ respectively.  
 
Rotation of this configuration by the matrices of the $SU(2)$ subgroup,  
which is defined by the third composite root $\vec \beta_3$ 
$$
U =  \left(\begin{array}{ccc}
\cos \frac{\theta}{2}&0&~\sin \frac{\theta}{2} e^{-i\phi}\\
0&1&0\\
-\sin \frac{\theta}{2} e^{i\phi}~&0&\cos \frac{\theta}{2}\\
\end{array}\right) \, , 
$$  
yields 
\begin{equation}  \label{two-higgs-rot}
U^{-1} \phi U =  \frac{1}{2}[v_1 + v_2] \tau^{(3)}_r +  
\frac{3}{4} [v_1 - v_2] {\tilde Q} \, .
\end{equation}
Up to the obvious reparameterization of the  
shape functions of the scalar field 
\begin{equation} \label{two-higgs}
\Phi_1 \to \frac{1}{2} \left[F_1(r)+F_2(r)\right],\qquad 
\Phi_2 \to \frac{\sqrt 3}{2} \left[F_1(r)-F_2(r)\right] \, ,   
\end{equation}
where the functions $F_1,F_2$ have the vacuum expectation values $v_1,v_2$, respectively, 
the configuration (\ref{two-higgs}) precisely corresponds to the third of the ans\"atze 
(\ref{h-1}). Because the $su(3)$-norm of the scalar field is set to be unity, 
the vacuum values must satisfy the condition  
$v_1^2 + v_2^2 - v_1v_2 =v $. 

Moreover, the reparameterization  (\ref{two-higgs}) allows us to write 
the scalar field of the $\beta_3$ monopole  
along positive direction of the $z$-axis as  
\begin{equation} \label{beta3-rep}       
\begin{split}
\vec \beta_3:~~~\phi(r\to \infty,\theta){\biggl| \biggr.}_{\theta=0} &= \left(v_1 - \frac{v_2}{2}
\right)H_1 +  
\frac{\sqrt 3}{2}v_2 H_2 \nonumber \\ 
&= (v_1 \vec \beta_1 + v_2 \vec \beta_2)\cdot \vec H = (\vec h \cdot \vec H) \ ,
\end{split}
\end{equation}
and we conclude that the asymptotic values  $v_1$ and $v_2$ 
are the coefficients of the expansion of the vector $\vec h$ in the basis of the simple 
roots and on the spatial asymptotic, the fields $F_1 (\vec \beta_1 \cdot \vec H)$ and 
$ F_2 (\vec \beta_2\cdot \vec H)$ can be identified with the  
Higgs fields of the first and second fundamental monopoles, respectively. 

Thus, the
embedding along the composite simple root $\vec \beta_3$ gives two fun\-da\-men\-tal 
mo\-no\-poles, which 
in the case of maximal symmetry breaking, are charged with respect to different $U(1)$ 
subgroups and are on top of each other. The configuration with minimal energy corresponds 
to the  boundary condition  $(\Phi_1)_{vac} =1, (\Phi_2)_{vac}=0$. 
We can interpret it by making use of 
Eq.~(\ref{two-higgs}), as two identical monopoles of the same mass. This degeneration is lifted as 
the value of the constant solution $\Phi_2=C$ increases, the vector of the Higgs field  $\vec h$ 
smoothly rotates in the root space and the boundary conditions begin to vary. 

According to the parameterization (\ref{two-higgs}), increasing of  
 $(\Phi_2)_{vac}$ results in the splitting of the vacuum values of the 
scalar fields of the first and second fundamental monopoles; the 
$\beta_1$-monopole is becoming heavier than the $\beta_2$-monopole. 
The maximal 
vacuum expectation value of the second component of the Higgs field of the $\beta_3$-monopole is 
$(\Phi_2)_{vac} = {1}/{2}$ or $v_2 = v_1/2$. 
This is a border value which 
separates the composite $\beta_3$-monopole from a single fundamental $\beta_i$-monopole, 
for which $(\vec h\cdot \vec \beta_i) \ge 0$ if $(\Phi_2)_{vac} \ge {1}/{2}$. 

\begin{figure}[t]
\begin{center}
\setlength{\unitlength}{1cm}
\begin{picture}(0,10.5)
\put(-4.1,10.6)
{\mbox{
\psfig{figure=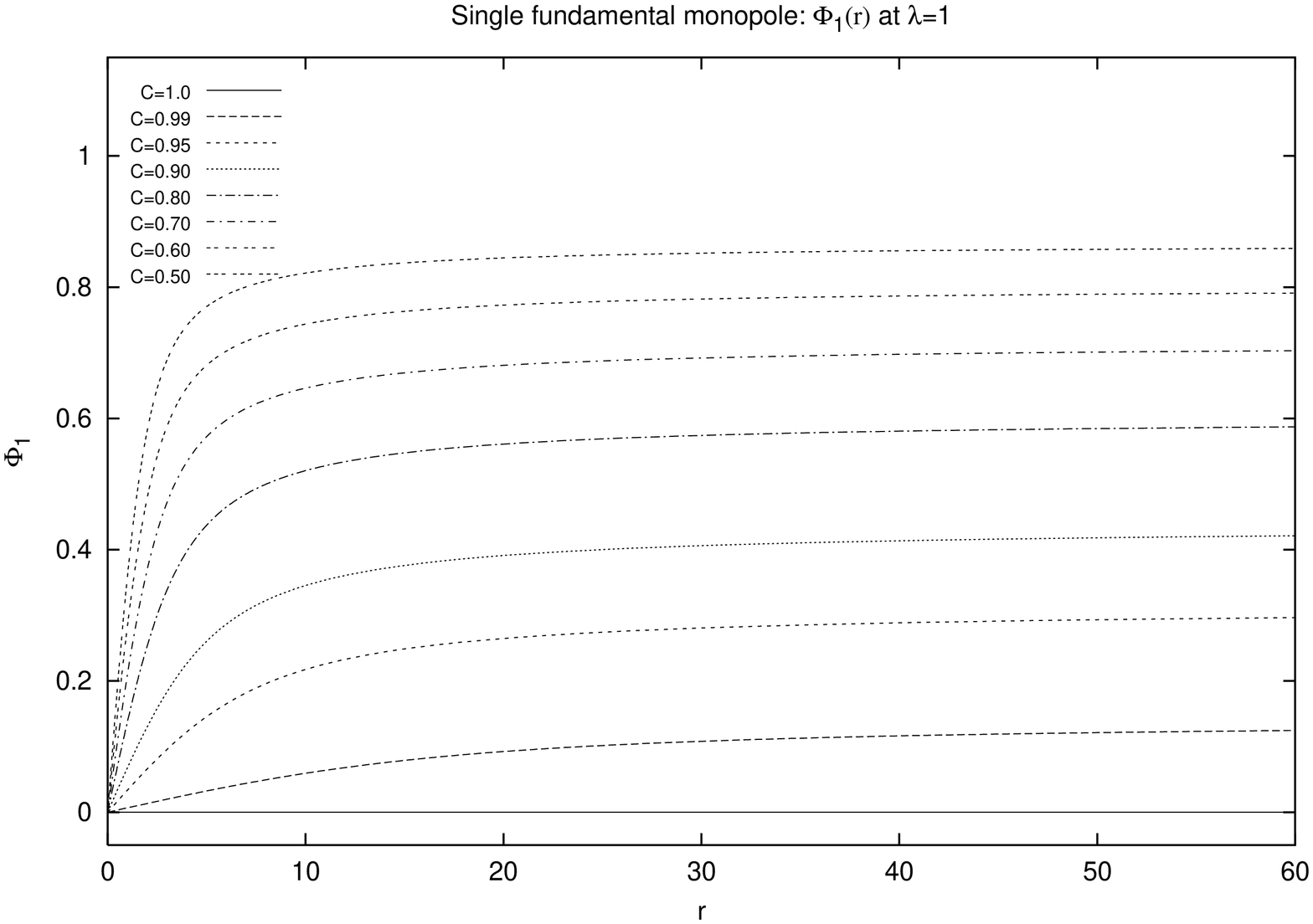,height=8.0cm, angle =-90}}} 
\end{picture} 
\begin{picture}(0,0.0)
\put(-4.2,4.9)
{\mbox{
\psfig{figure=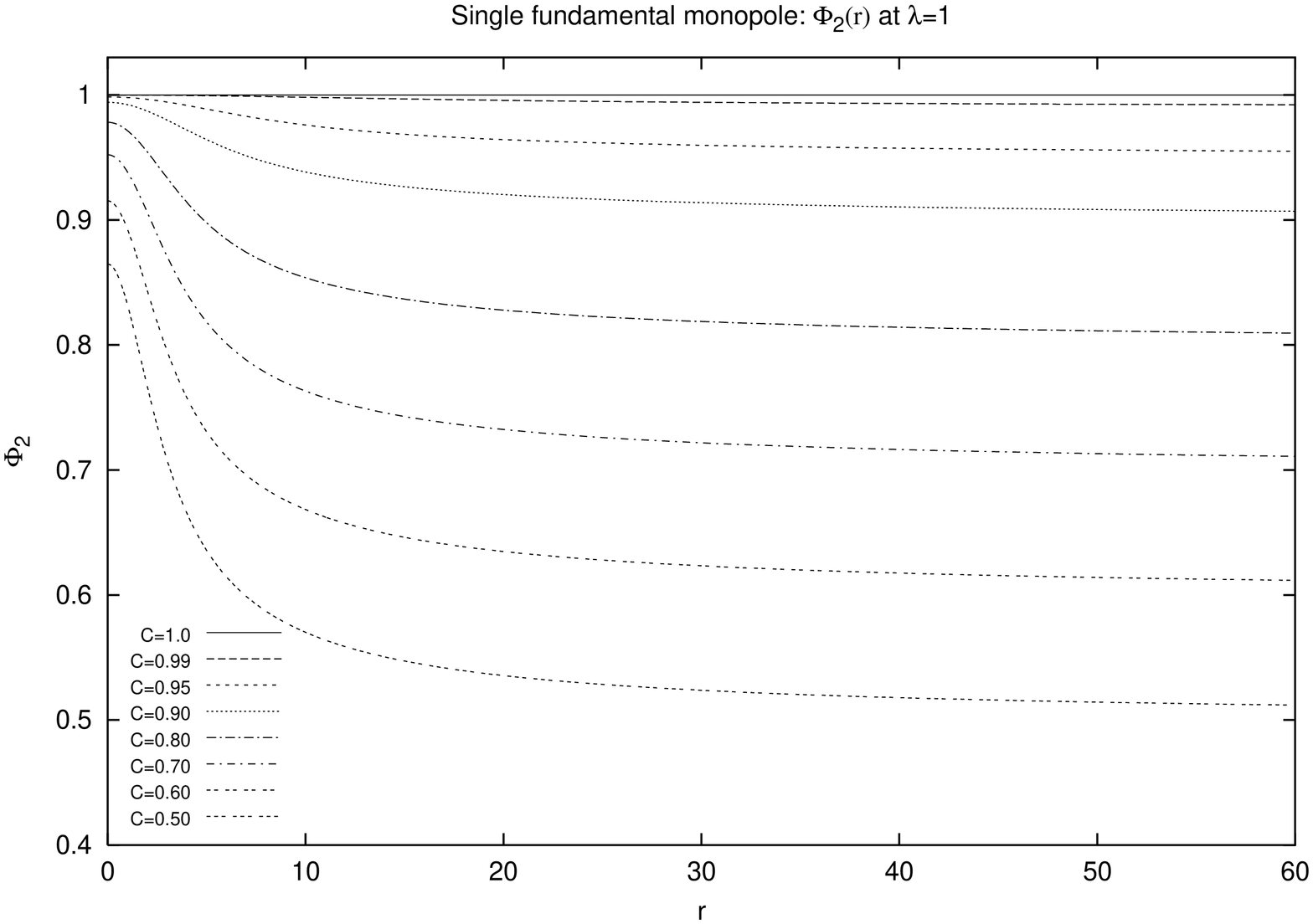,height=8.0cm, angle =-90}}} 
\end{picture} 
\end{center} 
\caption{
Structure functions of the Higgs field components $\Phi_1(r)$ and $\Phi_2(r)$ of the 
single fundamental monopole with different vacuum expectation values 
$(\Phi_2)_{vac}=C$ at $\lambda =1$.  
\label{fig8.6} }
\end{figure}

\begin{figure}[t]
\begin{center}
\setlength{\unitlength}{1cm}
\lbfig{f-ch8c}
\begin{picture}(0,7.0)
\put(-6.0,7.0)
{\mbox{
\psfig{figure=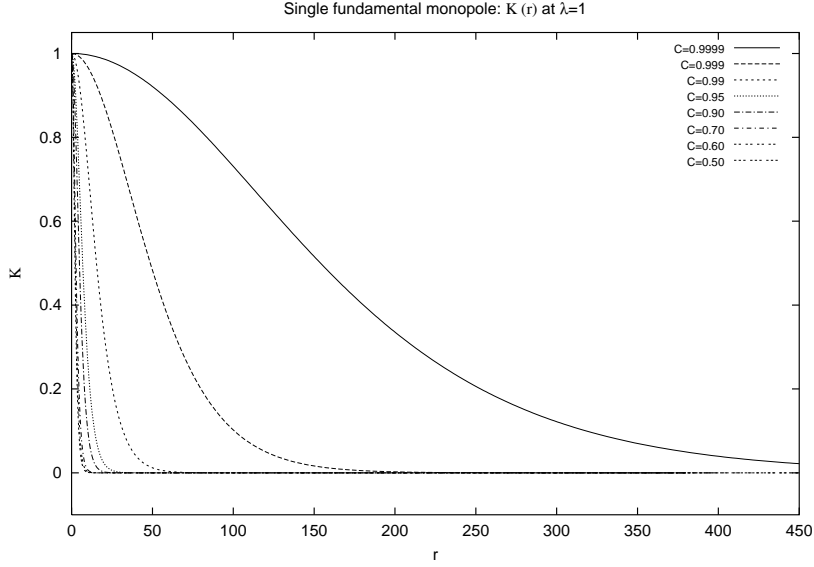,height=11.0cm, angle =-90}}} 
\end{picture} 
\end{center} 
\caption{
Structure function of the gauge field $K(r)$ of the 
single fundamental monopole for different vacuum expectation values 
$(\Phi_2)_{vac} = C$ at $\lambda =1$. \label{fig8.7} }
\end{figure}

As $(\Phi_2)_{vac}$ varies from $(\Phi_2)_{vac}=1/2$ to  $(\Phi_2)_{vac}=1$, the 
vector  $\vec h$ rotates clockwise for the $\beta_2$-monopole and anti-clockwise 
for the $\beta_1$-monopole within the same domain of the root space 
(Fig.~\ref{fig8.5}, left). The configuration smoothly moves to the limit  
$(\Phi_2)_{vac}=1$ when the vector  $\vec h$ becomes orthogonal to one of 
the simple roots. The numerical solution of the $SU(3)$ monopole shape functions 
is displayed in Figs.~\ref{fig8.6}, and \ref{fig8.7}.

Let us consider the behavior of a single fundamental monopole solution as 
the  vacuum expectation value $(\Phi_2)_{vac}$ approaches this limit \cite{Kunz93,S03}. 
Then, the ``hedgehog'' component $(\Phi_2)_{vac}$ tends to vanish and 
the monopole core spreads out as $(\Phi_2)_{vac}$ is approaches the limit $C=1$. 
This is the case of minimal symmetry breaking. 

\subsubsection{$SU(3)$ $\to$ $U(2)$:  Minimal Symmetry Breaking}

Let us analyze what happens if the scalar field becomes 
orthogonal to one of the simple roots. Suppose, for example, that ${\vec h}\cdot
\vec \beta_1 = 0$, that is, $\vec h = (0, 1)$. Then, as $r \to \infty$,
\begin{equation}                    \label{vacuum-phi-3}
\phi \to \phi_0 = v H_2 = \frac{v}{2 \sqrt 3}\left(\begin{array}{ccc}
1&0&0\\
0&1&0\\
0&0&-2\\
\end{array}\right) \, .
\end{equation}
As we have already mentioned, in this case, two eigenvalues of the mass matrix
coincide.  Clearly, the mass matrix then commutes with 
the generators of $SU(2)$ subalgebra $T^a_{(1)}$, which correspond to the $\vec \beta_1$ 
simple root and mixes the degenerated eigenvalues of $\phi_0$:
\begin{equation}
[\phi_0, T^a_{(1)}] = 0 \, .
\end{equation}
Furthermore, there is the $U(1)$ invariant subgroup. Indeed, the diagonal matrix $\phi_0$
(\ref{vacuum-phi-3}), as before, commutes with the electric charge operator $Q$.

Let us comment on the last statement. 
Recall that the electromagnetic subgroup of $SU(3)$ is not just
one of the Abelian subgroups generated by the elements of Cartan subalgebra
$\vec H$. The electric charge operator  $Q$  
is defined by Eq.~(\ref{Q-su}) as 
$Q = H_1 + H_2/{\sqrt 3} = \diag(2/3,-1/3,-1/3)$ 
and the
eigenvalues of the matrix $Q$ correspond to the electric charges of
the fundamental $SU(3)$ triplet (quarks).  Thus, the electromagnetic subgroup 
of spontaneously broken $SU(N)$ theory is compact. 
Indeed, the elements of this subgroup are given by 
$U = e^{i\alpha Q}$, and there are two points of the group manifold 
parameterized by the angles   
$\alpha$ and $\alpha + 2\pi N$, where $N=3$, which are
identical\footnote{There is a principal difference between the $SU(3)$
gauge theory and  the $SU(2) \times U(1)$ unified model of electroweak
interaction. In the latter, the electric charge operator is defined as 
a linear combination 
$$
Q_{ew} = \sin\theta_W  T^3 + \cos\theta_W Y, \qquad \sin^2\theta_W = 0.230 , 
$$
i.e., the electromagnetic subgroup of the Standard Model is
non-compact. Therefore, there is no topologically stable 
monopole solution within electroweak theory.}.

However it would be not correct to conclude that the 
invariant subgroup $H$ of the minimally broken $SU(3)$ model is a direct product
$SU(2) \times U(1)$. This is correct only with respect to the 
local structure of 
$H$, because the transformation of the electromagnetic $U(1)$ subgroup generated
by the electric charge operator contains the elements of the center $\mathbb{Z}_2 =
[-1,1]$ of the $SU(2)$ subgroup:
$$
e^{3\pi i Q} = \left(\begin{array}{ccc}
1&0&0\\
0&-1&0\\
0&0&-1\\
\end{array}\right)  \, .
$$
Hence, the group of residual symmetry of the vacuum is $H = SU(2) \times
U(1) /Z_2 \approx U(2)$, and there are two different classes of the
topologically non-trivial paths in $H$: the closed contours that encircle the
$U(1)$ subgroup of $H$ and the loops, which are traveling from the identity to
the element of center $\mathbb{Z}_2$ through the $U(1)$ subgroup, and back to the
identity through the $SU(2)$ subgroup \cite{Preskill84}. The monopoles with a
minimal $U(1)$ magnetic charge correspond to the contour travel only half-way
around the $U(1)$ subgroup, from the identity to the unit element of the center of
$SU(2)$. Such a monopole has a non-Abelian $\mathbb{Z}_2$ charge, as well as a non-Abelian
$SU(2)$ charge \cite{Preskill84}.

Thus, unlike (\ref{max-hom}), the second homotopy group for minimal symmetry breaking is
$$
\pi_2\left(\frac{SU(3)}{U(2)}\right) = \mathbb{Z} \, ,
$$
and there is only one topological charge. Indeed, for given orientation of the
Higgs field in the simple root basis, the topological charge, which is  
defined by formula (\ref{g-int-su}), becomes
$$
G =  {\vec g} \cdot {\vec h} = \frac{4 \pi }{e}
\left(n_1 {\vec \beta_1}^* + n_2 {\vec \beta_2}^*\right) 
\cdot {\vec h} =  n_2 \frac{4 \pi}{e} \frac{\sqrt 3}{2}\, .
$$

Thus, only the integer $n_2$, which corresponds to the non-orthogonal to the vector ${\vec h}$ 
simple root   $\vec \beta_2$,
is associated with the topological charge $G$  \cite{Taubes81}. 

As was pointed out by E.~Weinberg, one can understand the origin of this
reduction by taking into account the residual gauge freedom, which still 
exists within the chosen Cartan subalgebra ${\vec H}$. The point is that the
vector magnetic charge ${\vec g}$ is defined up to a transformation from the
Weyl subgroup, which does not take the vacuum $\phi_0$ out of the Cartan subgroup
\cite{Weinberg80}.  In the case of maximal $SU(3)$ symmetry breaking to $H =
U(1) \times U(1)$, there is just one fixed basis of simple roots given by the
vector $\vec h$ and both integers $n_1, n_2$ are topological charges
(\ref{wein-quant}). If the symmetry is broken minimally to $H = SU(2) \times
U(1)$, the condition that requires the inner product of the simple roots with
$\vec h$ to be positive, does not uniquely determine the basis of the
roots. There are two possible sets related by the Weyl reflection of the root
diagram that result from the global gauge transformations of the unbroken
$SU(2)$ subgroup. 
In the case under consideration, ${\vec h}$ is orthogonal to $\vec \beta_1$ and     
we can choose between two possibilities: $(\vec \beta_1, \vec \beta_2)$ and 
$(\vec \beta_1', \vec \beta_2') = (-\vec \beta_1, \vec \beta_1 + \vec \beta_2)$.

In the alternative self-dual basis, the vector magnetic charge reads  
\begin{equation}             \label{wein-quant-min}
{\vec g} = \frac{4\pi}{e} \left(n_1' 
{\vec \beta_1}^{\prime } + n_2' {\vec \beta_2}^{\prime }\right) \, ,
\end{equation}
where $n_1' = n_2 - n_1;~n_2' = n_2$. Therefore, only the invariant component of
the vector magnetic charge $\Tr ({\vec g} \cdot {\vec H} \phi_0) = 4\pi v n_2/e$, 
labeled by
the integer $n_2$, has a topological interpretation. Values of another integer
$n_1 = 0$ or $n_1 = 1$ are related to two possible 
orientations in the root space, which correspond to the two 
choices of the basis,  or the two possible ways
to embed the $SU(2)$ subgroup that we described above \cite{Weinberg80}.

\subsection{Massive and Massless Monopoles}
The question as to what could be the physical meaning of the components of the 
non-Abelian magnetic charge, which are abundant at non-maximal symmetry
breaking, has been discussed for years. We refer the reader to the
papers \cite{Bais98,Brand-Neri79,Murr89}.  Rather convincing seems to be the
argument that there is no topological restriction that forbids decay of
an arbitrary configuration into the state with minimal energy, that is, with a
minimal possible value of the non-Abelian magnetic charge
\cite{Coleman81,Brand-Neri79}.

However, in the BPS limit this argument is no longer valid, because all the
states, including the states with non-Abelian magnetic charge, correspond to
the same absolute minimum of energy. Thus, we have to understand what happens with 
a massive state in the limiting case of minimal symmetry breaking. 

Let us return to the $SU(3)$ model. 
We argued above that there are two different monopoles corresponding to the
two simple roots ${\vec \beta}_1$ and $ {\vec \beta}_2$, respectively.  However, it follows 
from the Bogomol'nyi bound (\ref{BPS-non-abel}) that only one configuration with a 
non-zero topological charge $n_2$ remains massive in the minimal symmetry breaking case
$(\vec h \cdot \vec \beta_1 = 0)$:
\begin{equation}
M = \frac{4 \pi v}{e}\left(n_1 
{\vec \beta}_1 \cdot  \vec h + n_2 \vec \beta_2 \cdot \vec h \right) =
M_2 = n_2 \frac{\sqrt 3}{2} \frac{4 \pi v }{e} \, . 
\end{equation}
Another monopole turns out to be massless: 
$$
M_1 = n_1 \frac{4 \pi v}{e} {\vec
\beta}_1 \cdot \vec h = 0 \, .
$$ 
This agrees with the results of our numerical calculations 
above; a single isolated fundamental monopole spreads out in space, 
its core radius increases as 
vector $\vec h$ approaches the limit where it becomes orthogonal to
one of the simple roots  (see Fig~\ref{fig8.7}). The mass of such an ``inflated'' 
monopole decreases  and tends to zero.  

However, the pattern of symmetry breaking becomes more complicated for a  
$\beta_3$-monopole, which is a composite state of two fundamental monopoles on top of 
each other \cite{S03}. Recall that  the idea is to treat 
minimal $SU(3)$ symmetry breaking as a special case of maximal symmetry
breaking, i.e., to analyse the rotation of the vector $\vec h$ in the root space. 
Indeed, it yields the  
splitting of the fundamental monopole masses as $(\Phi_2)_{vac}$ increases.  
One would expect that in the limiting case of minimal symmetry breaking, 
the $\beta_2$ monopole is becoming massless, that is, in that limit the vacuum value 
of the field $F_2$ should vanish, $v_2 \to 0$. However,  
two monopoles are overlapped and the presence of the massive monopole changes the situation.  
Indeed, the symmetry outside the core of the $\beta_1$-monopole is broken down to $U(1)$,  
which also changes the pattern of the symmetry breaking by the scalar field of the second monopole. 
One can see that the  vector  $\vec h$  becomes orthogonal to the simple 
root $\vec \beta_2$, when $(\Phi_1)_{vac} = {\sqrt 3}/{2}, (\Phi_2)_{vac} = {1}/{2}$, or 
$v_2 = {v_1}/{2}$. Going back to Eq.~(\ref{beta3-rep}), we can see that, in this case 
on the spatial asymptotic the scalar field  along the $z$-axis is  
$$
\phi(r\to \infty,\theta){\biggl| \biggr.}_{\theta=0} = \frac{3}{4} v_1 Q  \, ,
$$
where $Q$ is the electric charge matrix (\ref{Q-su}). Thus, the symmetry 
is still maximally broken and both monopoles are massive. 

Equation (\ref{beta3-rep}) indicates that the symmetry is minimally broken if 
the vector  $\vec h$  becomes orthogonal to the simple 
root $\vec \beta_1$  and $v_1 = {v_2}/{2}$. 
Then, the eigenvalues of the scalar field are 
the same as $H_2$, that is, the unbroken symmetry group is really 
$U(2)$. 
However, for the third composite root, such a situation corresponds to the negative value of the 
inner product $(\vec h\cdot \vec \beta_2)$ and it has to be excluded. Thus, the maximal 
value of the second component of the Higgs field of the $\beta_3$-monopole is 
$(\Phi_2)_{vac} = {1}/{2}$.

This conclusion allows us to understand what happens if we consider two distinct  
fundamental monopoles well-separated by a distance $R_0 \gg R_c$. As 
the vector 
$\vec h$ approaches to the direction orthogonal to either of the simple
roots, the core of the corresponding monopole tends to expand until its characteristic
size approaches the scale of $R_0$ \cite{Weinberg96,Chang}. At this stage, this monopole 
loses its identity as a localized field configuration. We have seen that, if this monopole 
were isolated, it would spread out and disappear, dissolving into the topologically trivial 
sector. However, as its core overlaps with the second massive monopole, it ceases to expand
\cite{Dancer92,Dancer93,Irwin97,Chang}.

Because at this stage the topological charge is resolved, this state is no longer a
topological soliton. E.~Weinberg \cite{Weinberg96,Weinberg98} suggested that
such a configuration be interpreted as a ``non-Abelian cloud'' of characteristic size
$R_0$, surrounding the massive monopole. The Coulomb magnetic field inside 
this cloud includes components that correspond to both  Abelian and
non-Abelian charges. However, on distances larger than $R_0$, only the Abelian component
is presented. The zero modes, which correspond to the massless monopole, are
transformed into the parameters of the non-Abelian global orientation and the
parameter characterizing the radii of these clouds \cite{Irwin97}.

\subsection{Interaction of Two Widely Separated $SU(3)$ Monopoles}

The conclusion that the multimonopole configurations appear in a rather natural way 
in a model with the gauge group rank greater than one caused a special
interest in the investigation of the low energy dynamics of these monopoles.
We have seen the the pattern of interaction between the monopoles is very different from a naive
picture of electromagnetic interaction of point-like charges 
\cite{Manton77,Manton-book,S-book,S05}.

Taubes pointed out \cite{Taubes81} that the short-range 
potential of interaction between the monopoles  
depends on the 
relative orientation of the monopoles in the group space, which is parametrised
by an angle $\delta$. There is, for example the  $SU(2)$ magnetic dipole solution 
\cite{mapKK,KKS} which corresponds to the 
saddle point configuration where the attractive short-range forces that are 
mediated by the $A_\mu^3$ vector boson and the Higgs boson, are balanced by 
the repulsion which is mediated by the massive vector bosons  
$A_\mu^\pm$ with opposite orientation 
in the group space. There is a difference from a system of two identically charged BPS 
monopoles where the scalar attraction is cancelled due to contribution 
of the repulsive gauge interaction.    

Let see how the situation changes for the  $SU(3)$ monopoles.  Again, we imply
a naive picture of the classical interaction of two fundamental monopoles, 
which are obtained by embedding of the single $SU(2)$ monopole 
configuration along the composite root ${\vec \beta}_1 + {\vec \beta}_2$. 
However, since these monopoles are charged with respect to different Abelian
subgroups, the character of interaction between the $SU(3)$ BPS  
monopoles depends on the type of embedding \cite{Weinberg96-2}.
This is also correct for the non-BPS extension \cite{S03}. 

Indeed, then there is only a long-range electromagnetic field that mediates 
the interaction between
two widely separated non-BPS monopoles, that is, they are considered as classical 
point-like particles with magnetic charges $g_i = \vec g_i \cdot \vec H = \vec \beta_i 
\cdot \vec H$. For a non-zero scalar coupling $\lambda$, the contribution of the scalar field 
is exponentially suppressed. 
The energy of the electromagnetic interaction then originates from 
the kinetic term  of the gauge field  
 $\sim \Tr\, F_{\mu\nu} F^{\mu\nu}$ in the Lagrangian
(\ref{Lagr}). Therefore,   
an additional factor $\Tr [(\vec \beta_i \cdot \vec H) (\vec \beta_j \cdot
\vec H) ] = (\vec \beta_i \cdot \vec \beta_j) $ appears in the formula for the energy
of electromagnetic interaction. In the case under consideration, $(\vec \beta_1 \cdot \vec \beta_2) 
= -{1}/{2}$, while $(\vec \beta_i \cdot \vec \beta_i) = 1$. 
This corresponds to 
the attraction of two different fundamental $SU(3)$ monopoles and repulsion of two 
monopoles of the same $SU(2)$ subalgebra due to the non-trivial group structure.  
The energy of interaction between the $\vec \beta_1$ and $\vec \beta_2$ monopoles is then: 
$$
V_{int} = - \frac{ ({\bf r}_1 {\bf r}_2)}{ r_1^3  r_2^3} \, .
$$ 

\begin{figure}   
\centering
\includegraphics[height=8.cm]{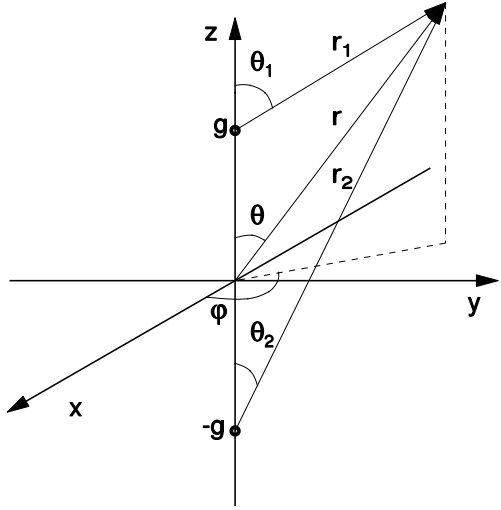}
\caption{
Electromagnetic field of two-monopole system.
\label{fig8.2a} }
\end{figure}

We can check this conclusion by making use of an analogy with the classical 
electrodynamics of point-like charges. Let us suppose that both monopoles are located 
on the $z$-axis at the points $(0,0,\pm R)$ (see Fig.~\ref{fig8.2a}).

The electromagnetic field of this configuration can be calculated in the Abelian gauge,  
where the gauge field becomes additive \cite{ArFreud}. If the monopoles are embedded along the same     
simple root, say  $\vec \beta_1$, we can write the components of the gauge field as 
\begin{equation}
A_r = A_\theta =0,\qquad A_\vphi = (1 + \cos \theta_1) \frac{\sigma_3^{(1)}}{2} + 
(1 + \cos \theta_2) \frac{\sigma_3^{(1)}}{2} \, .
\end{equation}
Simple calculation yields the components of the electromagnetic field strength tensor 
\begin{equation}  
\begin{split} 
F_{r\theta} &= 0;\qquad F_{r \vphi} = rR\sin^2\theta\left(\frac{1}{r_1^3}\, \frac{\sigma_3^{(1)}}{2}
-\frac{1}{r_2^3}\, \frac{\sigma_3^{(1)}}{2}\right)\, ,\\
 F_{\theta \vphi} &=-r^2\sin\theta\left(\frac{r-R\cos\theta}{r_1^3}\frac{\sigma_3^{(1)}}{2} + 
\frac{r+R\cos\theta}{r_2^3}\frac{\sigma_3^{(1)}}{2}\right)\, , 
\end{split}
\end{equation}
where $r_1$, $r_2$ are the distances of the point $r$ to the points at which monopoles are placed.
The field energy becomes 
\begin{equation}
\begin{split}
E = \Tr~ \left( \frac{1}{r^2 \sin^2\theta}  F_{r \vphi}^2 +  \frac{1}{r^4 \sin^2\theta}  
F_{\theta \vphi}^2\right) 
= \frac{1}{2}\left[ \left(\frac{{\bf r}_1}{r_1^3}\right)^2 + 
  \left(\frac{{\bf r}_2}{r_2^3}\right)^2 + \frac{2 ({\bf r}_1 {\bf r}_2)}{ r_1^3  r_2^3}
\right] \, ,
\end{split}
\end{equation}
that is, the potential energy of the electromagnetic interaction of two $\beta_1$ monopoles 
is repulsive. 
However, for a $\vec \beta_3$ configuration with vector charge $\vec g = (1,1)$, the components of the 
gauge fields are 
\begin{equation}
A_r = A_\theta =0;\qquad A_\vphi = (1 + \cos \theta_1) \frac{\sigma_3^{(1)}}{2} + 
(1 + \cos \theta_2) \frac{\sigma_3^{(2)}}{2} \, ,
\end{equation}
and, because $\Tr\, (\sigma_3^{(1)}\sigma_3^{(2)}) = - 1$, the field energy is 
\begin{equation}
E = \frac{1}{2}\left[ \left(\frac{{\bf r}_1}{r_1^3}\right)^2 + 
  \left(\frac{{\bf r}_2}{r_2^3}\right)^2 - \frac{ ({\bf r}_1 {\bf r}_2)}{ r_1^3  r_2^3}
\right] \, , 
\end{equation}
that is the $\vec \beta_1$ and $\vec \beta_2$ monopoles 
attract each other with a half-force compared to the case of the repulsion of two 
$\vec \beta_1$ 
monopoles.  

Let us now consider the BPS monopoles. 
To derive the metric on the $SU(3)$ moduli space, we shall analyze the situation in more 
detail \cite{MantonGibb95}.
Again, suppose that there are two fundamental monopoles at  
the points ${\bf r}_i$,~$i=1,2$, separated by a large distance $r$. 
The idea is to exploit an analogy with a classical
picture of the non-relativistic interaction between two point-like charges,
moving with small velocities ${\bf v}_i$. Thus, we consider a classic, 
long-range interaction between two particles with magnetic charges
 ${\vec g}_1, {\vec g}_2$ associated with the corresponding simple roots,  
\begin{equation}         \label{non-charge-su}
g_i =  (\vec g_i \cdot \vec H) =
\frac{4 \pi }{e} ({\vec \beta}_i \cdot {\vec H}) \, .  
\end{equation}
Because we considering the fundamental monopoles, the charge quantization
condition yields $g_1 = g_2 = g = 4\pi /e$.

Recall that some of the collective coordinates of the multi-monopole system
correspond to the dyonic degrees of freedom. These excitations shall transform
a monopole into a dyon.  
 Other collective
coordinates correspond to the spatial translations of the
configuration. We make an assumption that the electric charges of the dyons
are also vectors in the root space, that is \cite{Weinberg96-2}, 
\begin{equation}            \label{dyon-ch-non}
Q_i = q_i  ({\vec \beta}_i \cdot {\vec H})\, .   
\end{equation} 
Thus, the potential 
of interaction of two identical dyons remains proportional to the inner product $(\beta_i \cdot 
\beta_j)$, as in the case of a purely magnetically charged configuration.      

The canonical momentum of one of the dyons, which  is 
moving in the external field of another static dyon, is given by
\begin{equation}
{\bf P} = M {\bf v}_1 + q_1 {\bf A} + g {\bf {\widetilde A}} \, ,
\end{equation}
where the vector potentials of the electromagnetic field generated by the static
dyon are
$$
{\bf A} = g_2 {\bf a}\, , \qquad
{\bf {\widetilde A} } = - q_2 {\bf a} \, ,
$$
and we again make use of the definition ${\bf a}$ 
of the rescaled Dirac potential 
$$
[\pmb{\nabla} \times {\bf a}] = {\bf
r}/r^3 = -  \vecnabla (1/r)\, . 
$$  
Thus, far away from each of the dyons
their electric and magnetic fields are
\begin{eqnarray}
{\bf B}^{(i)} &=& g_i \frac{{\bf r} - {\bf r}_i}
{| {\bf r} - {\bf r}_i |^3} =
\frac{4 \pi}{e} (\vec \beta_i \cdot \vec H) 
\frac{{\bf r} - {\bf r}_i}
{| {\bf r} - {\bf r}_i |^3}\, ,\nonumber\\[2pt]
{\bf E}^{(i)} &=& Q_i \frac{{\bf r} - {\bf r}_i}
{| {\bf r} - {\bf r}_i |^3} =
q_i (\vec \beta_i \cdot \vec H) \frac{{\bf r} - {\bf r}_i}
{| {\bf r} - {\bf r}_i |^3} \, ,
\end{eqnarray}
which correspond to the scalar potentials
$$
A_0^{(i)} =  q_i\frac{ (\vec \beta_i \cdot \vec H)}
{| {\bf r} - {\bf r}_i |}\,  , \qquad {\widetilde A}_0^{(i)} = 
g \frac{(\vec \beta_i \cdot \vec H)}
{| {\bf r} - {\bf r}_i |} \, .
$$

In the evaluation of the energy of interaction between the two 
fundamental monopoles considered above, we take into account only 
the electromagnetic part and neglect the contribution of scalar fields. 
However, in the BPS limit, the Higgs field also becomes long-ranged
and the mass of the dyon is defined as
\begin{equation}             \label{mass-su}
M_i = \left(\vec h \cdot \vec\beta_i^*\right)\sqrt{q_i^2 + g^2} \, , 
\end{equation}
which is a generalization of (\ref{BPS-non-abel}). 
However, this formula is only correct for a single isolated dyon.

An external field of another dyon modifies the Coulomb-like tail of the scalar field as
\begin{equation}          \label{higgs-as-su}
\phi_i = 
v {\vec h} \cdot {\vec H} -  \sqrt{q_i^2 + g^2} \sqrt{1 - {\bf v}_i^2} ~
\frac{({\vec \beta}_i \cdot {\vec H})}{| {\bf r} - {\bf r}_i |} \, .
\end{equation}

As in the previous consideration of the interaction between two well separated $SU(2)$ 
monopoles we neglect here the difference between the masses of the monopole and the dyon, 
i.e., we assume for simplicity 
that both the velocities and the electric charges of the dyons are relatively small. 

We also have to take into account the Coulomb-like
potential associated with the scalar charge of the Higgs field, which is 
similar to the minimal $SU(2)$ model:
\begin{equation}        \label{dilat-su}
Q_D^{(i)} = ({\vec \beta}_i \cdot {\vec H}) \sqrt{q_i^2 + g^2} \, .
\end{equation}  
The long-range tail of the Higgs field yields some distortion of the
vacuum expectation value of the scalar field in the neighborhood of another monopole. 
As a result, the size of its core is increasing a bit,  
whereas the mass is decreasing:
\begin{equation}   \label{misalign}
M \to  
M - \frac{(\vec \beta_1^* \cdot \vec \beta_2^*)}{r} 
\sqrt{1 - {\bf v}_2^2} \sqrt{(q_1^2 + g^2)(q_2^2 + g^2)} \, .
\end{equation}

To sum up, the formal difference from the calculations we presented above consists in an additional factor $\lambda = -2 (\vec
\beta_1 \cdot \vec \beta_2)$, which appears as a coefficient at all terms of
interactions. Thus, by making use of simple analogy with Eq.~(\ref{L-2d-sep}),   
we can immediately write the Lagrangian of
relative motion of two widely separated $SU(3)$ BPS dyons of the same mass\footnote{
Of course, one may make use of the reduced mass $M = (M_1+M_2)/M_1M_2$, where 
$M_1,M_2$ are the masses of the dyons.} $M$ as
\cite{Weinberg96-2}: 
\begin{equation}
L = \frac{1}{4}\left(M + \frac{\lambda g^2}{r}\right) 
\left({\dot {\bf r}} \cdot 
{\dot {\bf r}} - \frac{Q^2}{g^2}\right) 
- \frac{\lambda}{2} gQ ~ {\dot {\bf r}} \cdot  {\bf a} \, , 
\end{equation}
where a relative charge of the pair of dyons is $Q = | q_1 - q_2 |$ and
the relative position is defined by the vector 
${\bf r} = {\bf r}_1 - {\bf r}_2$.

Recall that in the $SU(2)$ theory, the relative electric charge is connected with the gauge
cyclic collective coordinate $\Upsilon (t) $, which parameterizes the $U(1)$ subgroup. 
In the case of the $SU(3)$ model, the electric charges $q_1$ and $q_2$ are related to two 
different  $U(1)$ subgroups, which  can be
parameterized by two cyclic variables, the angles  
 $\alpha_1$ and $\alpha_2$, respectively. Then we may interpret the charges $q_1,q_2$ 
as conserved momenta conjugated to these variables. 
In the basis of a self-dual simple root that we are using, 
the period of the variables $\alpha_i$ is $T= 2 \pi e$ for both subgroups. 
The relative phase $\Upsilon = \alpha_1 - \alpha_2$ is a variable conjugated to the 
relative charge $Q$. 

The moduli space Lagrangian of relative motion must be
written in terms of the generalized velocities. 
Thus, by analogy with our discussion above we perform the Legendre transformation 
\index{Legendre transformation} 
$$
L({\bf r},  \Upsilon ) = L({\bf r}, Q) + 
gQ {\dot \Upsilon} \, ,
$$ 
where
\begin{equation}               \label{Q-su-1}
Q \equiv  \frac{2 g^3}{M + \frac{\lambda g^2}{r}}
\left(\dot \Upsilon +   (\vec
\beta_1 \cdot \vec \beta_2)\, ({\bf a} \cdot {\bf \dot r})\right) \, .
\end{equation}
Thus, we finally obtain the transformed Lagrangian of the relative motion
of two widely separated $SU(3)$ dyons \cite{MantonGibb95,Weinberg96-2}, 
\begin{equation}               \label{mod-lan-su}
L = \frac{1}{4}\left(M + \frac{\lambda g^2}{r}\right) 
{\dot {\bf r}} \cdot 
{\dot {\bf r}} + \frac{g^4}{M + \frac{\lambda g^2}{r}}
\left(\dot \Upsilon -\frac{\lambda}{2} ({\bf a} \cdot {\bf \dot r})\right)^2 \, .
\end{equation}

As before, this expression does not depend explicitly on the collective
coordinate $\Upsilon$. This means that the corresponding equation of motion
is just the condition of conservation of the relative electric charge
(\ref{Q-su-1}), rather than a dynamical equation. 

From this form of the Lagrangian (\ref{mod-lan-su}), we can read the asymptotic
metric for the moduli space of two widely separated $SU(3)$ BPS monopoles
\begin{equation}                 \label{metric-su3}
ds^2 = \left(1 + \frac{\lambda g^2}{M r}\right) d{\bf r}^2 +
\frac{\left( \frac{g^2}{2M}\right)^2} 
{1 + \frac{\lambda g^2}{Mr}}
\left(d \Upsilon +  (\vec
\beta_1 \cdot \vec \beta_2 )\, 
 ({\bf a} \cdot d{\bf r})\right)^2 \, ,
\end{equation}
that is, the metric of Taub-NUT space with the length parameter $\lambda g^2/M$. 

The principal difference from the asymptotic metric, which describes 
two $SU(2)$ dyons, is that the parameter 
$\lambda$ remains negative only if both of these fundamental
monopoles correspond to the same simple root.  We can expect this, because in
this case, there are two widely separated $SU(2)$ monopoles, and the 
corresponding  moduli space is described by the singular 
Taub-NUT metric (\ref{NUT-1}).  However, we have seen that, 
if these two monopoles correspond to the different simple roots, 
$\vec\beta_1$ and $\vec\beta_2$, we have \cite{Weinberg96-2}:
$$
\lambda =  
-2 (\vec \beta_1 \cdot \vec \beta_2) = -2 [(1,0) \cdot 
(-1/2, {\sqrt 3}/2)] = 1\, .
$$ 

Note that the conservation of both the total and relative electric charges 
results from the 
conservation of the individual electric charges $q_1,q_2$ of each monopole. This yields
a $U(1)$ symmetry of the metric of relative motion, which is not a symmetry of the 
Atiyah--Hitchin metric, where relative charge, in general, is not an integral of motion. 
This symmetry simplifies the low-energy dynamics of two distinct $SU(3)$ monopoles; there is 
no right-angle scattering, but bounce trajectory in a head-on collision.  

\section{$N=2$ Supersymmetric $SU(2)$ Magnetic Monopoles}

One of the reasons why nowadays there is such a strong interest in the
monopole problem is a subject that we have not discussed yet, 
namely the role these configurations playing in supersymmetric theories.  
This is a very remarkable subject since, for example, $N=2$ $SU(2)$ Yang-Mills 
model can be considered as a supersymmetric cousin of the Georgi--Glashow theory. 
As we shall see below, these 
models yield a very nice frame to reveal most of the remarkable features of the 
monopoles from a single point point of view. 
Indeed, a supersymmetrical theory by very definition includes both bosonic and 
fermionic variables. Thus, there is a nice natural 
interplay between the localized monopole 
configuration and the fermions in such a model. Futhermore, only BPS monopoles 
appear as non-perturbative solutions of the supersymmetric theory. 

The elegant and beautiful structure of the $N=2$ SUSY Yang--Mills theory
has very little difference from the conventional model, but there is 
the miracle of supersymmetry, which makes the model exactly solvable. 
The hope of 
theoreticians is that there be some properties of the SUSY Yang--Mills theory 
that in the non-perturbative regime can shed a new light on the problem of QCD 
confinement. Therefore, it looks very interesting that the monopole-like configuration 
plays a crucial role in the low-energy supersymmetric dynamics. Moreover,
the conjecture of electromagnetic duality becomes an exact property of
such a theory, where the gauge and the monopole sectors are related to each other by 
some transformation of duality. A clue to the latter solution is the  
observation that an exact duality on a quantum level could only survive if the 
Bogomol'nyi bound is not violated by the quantum corrections. This becomes 
possible if the bosonic and fermionic loop diagrams, which could affect the 
effective potential, are mutually cancelling. This mechanism of cancellation
is a familiar property of supersymmetric models.

In our consideration we shall not discuss the technical aspects of 
supersymmetry. There are many 
good reviews on supersymmetry that give a thorough introduction into the 
basis of supersymmetry. We refer the reader, for example,  
to \cite{Wess,Bilal00,Lykken96,Sonius85} and references therein.
Here, we only briefly comment on the properties of the supersymmetric monopoles. 

\subsection{Praise of Beauty of $N=2$ SUSY Yang--Mills}
The Lagrangian of pure Yang-Mills $N=2$ SUSY can be written as 
\bea     \label{L-N=2-Dirac}
L^{N=2} &=& -\frac{1}{4} F^{\mu\nu}_a F_{\mu\nu}^a + 
\frac{\theta e^2}{32 \pi^2} F^{\mu\nu}_a {\widetilde F}_{\mu \nu}^a
+ \frac{1}{2}(\nabla_\mu \phi^a_1)^2 \\
&+& \frac{1}{2}(\nabla_\mu \phi^a_2)^2 - \frac{e^2}{2} 
[\phi^a_1, \phi^b_2]^2 - 
i{\bar \chi} \gamma^\mu \nabla_\mu \chi - e \phi_1{\bar \chi}\chi  
- ie \gamma_5 
\phi_2 {\bar \chi}\chi \, . \nonumber
\eea

Thus, this is a Lagrangian of the system of interacting bosonic and fermionic 
fields that is very similar to the Georgi--Glashow  
model coupled with fermions. The condition 
of supersymmetry only imposes the restriction on the masses of the particles  
of the same multiplet and requires the coupling constant of the vector 
and scalar multiplets to be unique. 
So, there is nothing strange in the model
(\ref{L-N=2-Dirac}) and, as noted in short review \cite{Flume96}, 
what really looks unusual there,  
is the perfection and the universality of such a model, which actually  
incorporates almost all non-trivial 
elements of the modern quantum field theory and, therefore, deserves to be used as a 
{\it ``Model to Teach Quantum Gauge Theory''}. The properties of this very  
remarkable theory include the following. 
\begin{itemize}
\item[$\bullet$] It describes non-Abelian gauge fields coupled with the matter fields.
\item[$\bullet$] It includes scalar and pseudoscalar fields with minimal Yukawa couplings 
to the fermions.
\item[$\bullet$] It is renormalizable.
\item[$\bullet$] It is asymptotically free.
\item[$\bullet$] The Higgs mechanism is used to generate the masses of 
the particles softly.
\item[$\bullet$] The non-perturbative sector of the model contains both monopoles and instantons.
\item[$\bullet$] There is a chiral symmetry between scalar and pseudoscalar fields of 
the classical Lagrangian. However, on a quantum level 
the anomaly breaks this symmetry to the discrete subgroup $\mathbb{Z}_4$. 
\item[$\bullet$] The corresponding algebra of SUSY contains the central charges.
\item[$\bullet$] The action of the model is scale-invariant but there is a quantum 
anomaly in the trace of the energy-momentum tensor.  
\item[$\bullet$] The model provides a realization of the Montonen--Olive conjecture of duality.
\item[$\bullet$] The model arises from the string theory in the point-particle limit.
\end{itemize} 

Let us choose $SU(2)$ as the gauge group of the model. It makes it even 
closer to the Georgi--Glashow model. The differences now are in the 
structure of the scalar potential $U[\phi]$, appearance of two real fields 
$\phi_1, \phi_2$ 
which can be considered as components of a complex scalar field $\phi^a = 
\frac{1}{\sqrt 2}(\phi_1^a + i \phi_2^a) $, and that all the fields, including the fermions,
take values in the adjoint representation of the gauge group.
However, the main difference is that, unlike the genuine Georgi--Glashow model (\ref{Lagr}),
 the potential of the scalar field   
$$
U = \frac{e^2}{2}(f_{abc} \phi^{b^\dagger} \phi^c)^2
$$
does not fix the Higgs vacuum uniquely . 
Indeed, this potential vanishes if $\phi$ 
and $\phi^\dagger$ commute with each other. 

For a given choice of the 
$SU(2)$ gauge group, this condition is satisfied if these fields take values 
in the Abelian subgroup of $SU(2)$: 
\begin{equation}        \label{vacuum-scalar-SUSY}
\phi_0 =  v T^3 =  v\frac{ \sigma^3}{2} \, ,
\end{equation}
where $v$ is now an arbitrary complex number. More generally, 
for a model with the gauge group $G$, the Higgs vacuum is defined by the relation 
$$
[\phi_0,\phi_0^\dagger] = \varepsilon_{abc}\phi^{b\dagger}\phi^c = 0 \, , 
$$ 
which means that $\phi_0$ lies in the diagonal 
Cartan  subalgebra $\vec H = (H_1, H_2 \dots H_{N-1})$ of $G$.  

Let us note that we already considered a similar definition of the Higgs vacuum  
when we discussed 
the properties of the $SU(N)$ monopoles 
(cf.~Eq.~(\ref{vacuum-higgs-wein}) and the following discussion above). 
Thus, the Higgs vacuum breaks the gauge group to 
$H$ and there is a set of gauge inequivalent classical vacua of the $N=2$ SUSY 
Yang--Mills theory. However, the $N=2$ SUSY is still in force and it remains 
a symmetry of these vacua.

The very important point is that each of these vacua corresponds to a  
different physical situation. Indeed, we know, for example, 
that the vacuum value of the 
scalar field generates the masses of particles via the Higgs mechanism and 
the physical observables are, therefore, directly related  with $\phi_0$.
Thus, unlike the simple Georgi--Glashow model, 
there is a space of different vacuum values of the scalar field of dimension 
equal to the rank $r$ of the gauge group $G$.  
Within the framework of the full quantum theory the moduli space is parameterized
by the vacuum expectation values of the Higgs field $<\phi>$ \cite{WS}. 
However, we shall not discuss quantum aspects of the monopole theory here. 
Our primary subject
is the spectrum of the monopole solutions 
of the classical $N=2$ SUSY pure Yang--Mills theory. 

\subsection{Construction of $N=2$ Supersymmetric $SU(2)$ Monopoles}
Let us show now that there are monopole solutions of the model with the action 
(\ref{L-N=2-Dirac}). As before, in the case of the `t Hooft--Polyakov 
monopole solution of the Georgi--Glashow model, we can consider the static, time-independent 
configurations first, and ignore the topological $\theta$-term for a while.

Recall that the difference of the bosonic sector of the supersymmetric 
action (\ref{L-N=2-Dirac})  from the Georgi--Glashow model is that there are two 
real scalar fields, $\phi_1$ and $\phi_2$, which correspond to the scalar and
pseudoscalar coupling to the Dirac fermions. For the sake of simplicity, let us suppose 
that asymptotically the complex scalar field lies on the unit sphere, that is,  
$| \phi | = 1$ as $r \to \infty$. 
 
Then, the bosonic 
Hamiltonian of the static $SU(2)$ super Yang--Mills theory 
can be written in the form 

{\small
\begin{equation}                            \label{En-lower-N=2}
\begin{split}
E &=\frac{1}{2}\int d^3 x \biggl\{ (E_n^a )^2 + 
(B_n^a)^2 + (\nabla_n \phi^a_1)^2  
+ (\nabla_n \phi^a_2)^2 
+ e^2 (\varepsilon_{abc} \phi^b_1\phi^c_2)^2 \biggr\} \nonumber\\[4pt] 
&=\frac{1}{2} \int\! d^3 x  \biggl\{ 
\left[E_n^a \!-\! \nabla_n \phi^a_1 \sin \delta
\!-\! \nabla_n \phi^a_2 \cos \delta \right]^2
+  \left[B_n^a \!-\! \nabla_n \phi^a_1 \cos \delta 
\!+\! \nabla_n\phi^a_2 \sin \delta \right]^2\biggr\} \nonumber\\[4pt]
&+ 2\int d^3 x \biggl\{ E_n^a (\nabla_n \phi^a_1 
\sin \delta + 
\nabla_n \phi^a_2 \cos \delta) +  B_n^a (\nabla_n \phi^a_1 
\cos \delta - \nabla_n \phi^a_2 \sin \delta)\biggr\}\nonumber\\[4pt]
&+\frac{e^2}{2}\int d^3 x ~(\varepsilon_{abc} \phi^b_1\phi^c_2)^2 \, , 
\end{split}
\end{equation}

}
\noindent where $\delta$ is an arbitrary parameter and
the symbol $\nabla_n$ denotes a covariant derivative.  

Thus, if the scalar potential is vanishing, the lower energy bound is given by
the system of equations
\bea        \label{BPS-eq-SUSY}
E_n^a &=&  \nabla_n\phi^a_1 \sin \delta
+ \nabla_n\phi^a_2 \cos \delta \equiv  \nabla_n {\tilde \phi}^a_1\, ,\nonumber\\  
B_n^a &=& \nabla_n \phi^a_1 \cos \delta - \nabla_n\phi^a_2 \sin \delta 
\equiv \nabla_n {\tilde \phi}^a_2 \, ,
\eea
where we define the linear combinations of the scalar fields 
\begin{equation} \label{rot-higgs}
{\tilde \phi}^a_1 = \phi^a_1 \sin \delta + \phi^a_2 \cos \delta,
\qquad  {\tilde \phi}^a_2 = \phi^a_1 \cos \delta - 
\phi^a_2 \sin \delta \, ,
\end{equation}
which appear in this supersymmetric counterpart of the BPS equations (\ref{BPS}).
On the other hand, a proper 
parameterization of the vacuum manifold of the model  
is still given in terms of the asymptotic values of the fields $\phi_i$, rather than 
the rotated fields ${\tilde \phi}_i$, because they are invariant with respect to the 
transformations of the modular group $SL(2,\mathbb{Z})$. 

A particular representation of these two scalar fields as  
$$
\phi^a_1 = \phi^a \cos \delta,\qquad \phi^a_2 = -\phi^a \sin \delta \, , 
$$
immediately yields the BPS equations  (\ref{BPS}) for the field ${\tilde \phi}_1 =0,~~  
{\tilde \phi}_2 = \phi$, 
up to an obvious identification of the free angular parameters. Thus, the
general static spherically symmetric monopole solution in the bosonic sector may be 
obtained as a  generalization of the BPS solution:
\bea                                 \label{BPS-Sol}
\phi^a_1 = \frac{r^a}{e r^2} H (r)\cos \delta, 
&\quad&\phi^a_2 = -\frac{r^a}{e r^2} H (r)\sin \delta,\nonumber\\ 
A_n^a =  \varepsilon _{amn}\frac{r^m }{ er^2} (1 - K (r))\, ,
\eea
where the structure functions are the well-known analytical solutions
(\ref{BPS-solu}), which are expressed via the dimensionless parameter $\xi = er$:
\begin{equation}                      \label{BPS-solu-11}
K = \frac{\xi}{\sinh \xi }, \qquad H = \xi \coth \xi - 1 \, .
\end{equation} 
Asymptotic behavior of the complex Higgs field is now   
\begin{equation} 
\phi^a \to e^{-i\delta} {\hat r}^a\left(1 - \frac{1}
{er}\right) 
\quad {\rm as}\quad \xi 
\rightarrow \infty \, .
\end{equation} 
Let us note that the real components, $\phi_1$ and $\phi_2$, may have different
behavior while they are approaching this regime.

It is natural to define two-component electric and magnetic charges as 
\cite{Fraser97,Gauntlett00,Gauntlett01}:

\begin{equation}     \label{SUSY-q-g}
q^i = \int d^3x~ \partial_n \left(E_n^a \phi^a_i\right),\qquad 
g^i = \int d^3x~ \partial_n \left(B_n^a \phi^a_i\right) \, .
\end{equation} 
Then the bound on the mass of $N=2$ supersymmetric monopoles becomes
\begin{equation}    \label{BPS-delta-rot}
M \ge [(g^1 + q^2) \cos \delta + (q^1 - g^2) \sin \delta] \, ,
\end{equation}
which implies that the Bogomol'nyi bound is saturated if 
\begin{equation}     \label{BPS-constr}
\tan \delta = \frac{q^1 - g^2 }{g^1 + q^2} \, , 
\end{equation}
and then 
$E_n^a = \nabla_n {\tilde \phi}^a_1$,~~ $B_n^a = \nabla_n {\tilde \phi}^a_2$.  

So far, we brutally set ${\tilde \phi}_1$ to be identically zero and 
make use of the straightforward analogy with construction of the BPS monopoles 
in the first section. However, one can construct a wider class of monopoles 
by relaxing this restriction. Indeed, 
the Gauss law (\ref{Gauss})  for such a static field configuration with a two-component 
scalar field takes the form
$$
\nabla_n E_n - ie [ {\tilde \phi}_1, \nabla_0  {\tilde \phi}_1] - ie  
[ {\tilde \phi}_2, \nabla_0  {\tilde \phi}_2] =0 \, .
$$
Let us now impose a gauge $A_0 =   {\tilde \phi}_1$. Then the Gauss law  
becomes a covariant Laplace equation for the field $ {\tilde \phi}_1$:
\begin{equation}   \label{second-BPS}
\nabla^2 {\tilde \phi}_1 - e^2 [ {\tilde \phi}_2, [ {\tilde \phi}_1, {\tilde \phi}_2]] = 0 \, ,
\end{equation}
which is referred to as the {\it ``secondary BPS equation'' } 
\cite{Gauntlett00,Gauntlett01,Lee98,Tong99}. Certainly, this equation becomes even more 
restrictive if we only consider the configurations that satisfy 
$[ {\tilde \phi}_2, {\tilde \phi}_1] = 0$.

The {\it ``primary BPS equation''} is obviously $B_n = \nabla_n {\tilde \phi}_2$.  
For a given solution of this equation, 
the secondary BPS equation describes a large gauge transformation of   
the fields of the BPS monopole $A_k, {\tilde \phi}_2$ \cite{Lee98}.  
Thus, the solution of the secondary BPS equation yields the gauge zero modes about the 
original monopole configuration. This zero mode exists for each solution of the primary BPS 
equation. We already know that these gauge transformations 
generate an electric charge of the configuration that transforms a monopole into a dyon.

Taking into account the Bianchi identity for the 
magnetic field and the equation of motion for the electric field, 
we can now recover the  lower bound on the mass of the dyon:
\cite{Fraser97,Lee98}: 
\begin{equation}                   \label{BPS-bound-11}
M \ge \mid (q^1 - g^2) +  i (g^1 + q^2)\mid \, .
\end{equation}

So far, we are discussing the model with the gauge group $SU(2)$. In this case, 
the vanishing of the scalar potential, $[\phi_1,\phi_2]^2 = 0$, means that 
both $\phi_1$ and 
$\phi_2$ belong to the unique Cartan subgroup of the gauge group. In other
words, $\phi_1$ must be proportional to $\phi_2$ and, therefore, the components 
of the electric and magnetic  charge vectors are also proportional to each other. 
Thus, $g^1q^2 = g^2q^1$ and  
the bound (\ref{BPS-bound})  is recovered from (\ref{BPS-bound-11}):
$$
M \ge \sum_{i=1,2} \sqrt{q_i^2 + g_i^2} \, .
$$
However, the principal difference from the Georgi--Glashow model is that 
now the BPS bound is directly related with the algebra of supersymmetry. 
Moreover, the electric and the magnetic charges (\ref{SUSY-q-g}) are the real and 
the imaginary components of the $N=2$ central charge $Z$, respectively \cite{Witten78}.

\subsection{Central Charges in the $N=2$ SUSY Yang--Mills}
Supersymmetry of the model (\ref{L-N=2-Dirac}) means that there 
are $N=2$ anticommuting generators  $Q^1_\alpha$,$Q^2_\alpha$, which transform 
as spin-half operators in the  $(\frac{1}{2},0)$ and 
$(0,\frac{1}{2})$ representations of the Lorentz group, respectively. The
corresponding  $N=2$ SUSY algebra  
includes a complex central charge $Z$:
\be     \label{z-algebra}
 \{Q^1_\alpha,Q^2_\beta\} = 2\varepsilon_{\alpha\beta}Z \, .
\ee

Note that the generators of $N=2$ 
supersymmetry $Q^1_\alpha$,$Q^2_\alpha$ by definition are the charges  
which correponds to the supercurrents $S_\mu$, $\bar S_\mu$. These currents 
appear as a result of the infinitesimal supersymmetric 
variations of the component fields with the 
parameter of supertranslation
$\xi$: 
\begin{equation}         \label{N=2-Dirac-trans}
\begin{split}
\delta \chi^a &= \biggl(\sigma^{\mu\nu} F^a_{\mu\nu} + 
\gamma^\mu~\nabla_\mu(\phi_1^a + \gamma_5\phi_2^a)\biggr)\xi\, ,\\ 
 \delta A^a_\mu &= i{\bar \xi} \gamma_\mu \chi^a - i{\bar \chi}^a \gamma_\mu \xi\, ,\\
\delta \phi_1^a &= i{\bar \xi} \chi^a - i{\bar \chi}^a \xi\, ,\\
\delta \phi_2^a &= i{\bar \xi} \gamma_5 \chi^a - i{\bar \chi}^a \gamma_5 \xi
 \, . \end{split}
\end{equation}
It is instructive to apply these transformations to see that they 
change the $N=2$ Lagrangian (\ref{L-N=2-Dirac}) by a total derivative. 
Explicitly the corresponding supercurrents are, 
\begin{equation}
\begin{split}        \label{spinor-comp-supercurrent}
{S_\mu^{(1)}}_\alpha &= 
 -(\sigma^\nu)_{\alpha{\da}} {\bar \lambda}^{a{\da}}
(iF_{\mu\nu}^a + {\widetilde F}_{\mu\nu}^a) 
+ {\sqrt 2} (\sigma_\nu {\bar \sigma}_\mu \psi^a)_\alpha \nabla^\nu 
\phi^{a\dagger}  + (\sigma_\mu)_{\alpha {\da}} {\bar \lambda}^{a{\da}}
\phi^\dagger T^a \phi,\\[4pt]
{S_\mu^{(2)}}_\alpha &= -(\sigma^\nu)_{\alpha{\da}} {\bar \psi}^{a{\da}}
(iF_{\mu\nu}^a + {\widetilde F}_{\mu\nu}^a) 
- {\sqrt 2} (\sigma_\nu {\bar \sigma}_\mu \lambda^a)_\alpha \nabla^\nu 
\phi^{a\dagger}  + (\sigma_\mu)_{\alpha {\da}} {\bar \psi}^{a{\da}}
\phi^\dagger T^a \phi .
\end{split}
\end{equation}
Here the bispinor field $\chi$, which arrears in (\ref{L-N=2-Dirac}) is decomposed into 
two Weyl spinors $\lambda$ and $\psi$ as
$$
\chi = \begin{pmatrix}\psi_\alpha \\ -i{\bar \lambda}^{\da} \end{pmatrix},
\qquad 
{\bar \chi} = (i\lambda^\alpha, {\bar \psi}_{\da}) \, ,
$$ 
and 
we 
make use of the property $\varepsilon \sigma^\mu {\bar \psi} = - {\bar \psi} 
{\bar \sigma}^\mu \varepsilon$ and the following identities involving the Pauli matrices 
\begin{equation}
\begin{split} 
\sigma^\mu {\bar \sigma}^\nu \sigma^\rho &= g^{\mu\nu} \sigma^\rho 
- g^{\mu\rho} \sigma^\nu + g^{\nu\rho} \sigma^\mu + 
i\varepsilon_{\mu\nu\rho\omega}\sigma^\omega \, ,\\
{\bar \sigma}^\mu \sigma^\nu {\bar \sigma}^\rho &= g^{\mu\nu} 
{\bar \sigma}^\rho 
- g^{\mu\rho} {\bar \sigma}^\nu + g^{\nu\rho} {\bar \sigma}^\mu - 
i\varepsilon_{\mu\nu\rho\omega}{\bar \sigma}^\omega \, .
\end{split}
\end{equation} 
Therefore, the anticommutator (\ref{z-algebra}),  
which yields the central charge, can be written as the equal-time anticommutator
of the volume integrals over the temporal components of these currents: 
\begin{equation}         \label{comm-supersurrents}
\{Q^1_\alpha,Q^2_\beta\} = \left\{\int d^3x~  S^{(1)}_{0 \alpha}({\bf x}),
 \int d^3x'~  S^{(2)}_{0 \beta}({\bf x'})\right\} = 
2\varepsilon_{\alpha\beta}Z \, .
\end{equation}
Olive and Witten noticed \cite{Witten78} that this commutator non-vanishes
due to the boundary terms, which are just the electric and the magnetic charges of 
the configuration. Indeed, let us note that ${\bar \lambda}_{\da} = \lambda_\alpha^\dagger$ and
$$
{\bar \lambda}^{\da} = \varepsilon^{{\da}{\db}} {\bar \lambda}_{\db} = 
i(\sigma_2 \lambda^\dagger)^{\da} \, .
$$
It is clear that the 
temporal components of the supercurrents  (\ref{spinor-comp-supercurrent}) are 
\be
\begin{split}
S_{0\alpha}^{(1)}&=-i(\sigma_k \sigma_2 \lambda^{a\dagger})_\alpha (i F^a_{0k} + 
{\widetilde F}^a_{0k}) 
+ {\sqrt 2} (\sigma_k \psi^a)_\alpha \nabla_k \phi^{a\dagger} + i(\sigma_2 \lambda^{a\dagger})_\alpha 
\phi^\dagger T^a \phi,\nonumber\\[4pt]
S_{0\alpha}^{(2)} &=-i(\sigma_k \sigma_2 \psi^{a\dagger})_\alpha (i F^a_{0k} + 
{\widetilde F}^a_{0k}) 
- {\sqrt 2} (\sigma_k \lambda^a)_\alpha \nabla_k \phi^{a\dagger} + 
i(\sigma_2 \psi^{a\dagger})_\alpha 
\phi^\dagger T^a \phi \, . \nonumber\\
\end{split}
\ee
To evaluate the central charge $Z$, we substitute 
these expressions into  anticommutator (\ref{comm-supersurrents}) and make use 
of the anticommutation relations for the fields $\psi$ and $\lambda$. The 
relevant terms in  Eq.~(\ref{comm-supersurrents}) are 
\be
\{Q^1_\alpha,Q^2_\beta\} = i {\sqrt 2}\int d^3x \left[(\sigma_i \sigma_2 \sigma_j^T)_{\alpha \beta} -   
(\sigma_i \sigma_2 \sigma_j^T)_{\beta\alpha}\right](iF^a_{0k} + {\widetilde F}^a_{0k}) \nabla_k \phi^{a\dagger}\, .
\ee
Next, making use of the algebra of the Pauli matrices 
$$(\sigma_i \sigma_2 \sigma_j^T)_{\alpha \beta} = \left(\sigma_2 (-\delta_{ij} 
+ i\varepsilon^{ijk}\sigma_k^T)\right)_{\alpha\beta} \, ,
$$ we obtain  
\bea
\{Q^1_\alpha,Q^2_\beta\} &=&  -2 i {\sqrt 2}\int d^3x~(\sigma_2)_{\alpha\beta} \delta_{ij}~( 
iF^a_{0k} + {\widetilde F}^a_{0k})  \nabla_k \phi^{a\dagger}\nonumber\\
&=&-2 {\sqrt 2}\varepsilon_{\alpha\beta} 
\int d^3x~ (iF^a_{0k} + {\widetilde F}^a_{0k}) \nabla_k \phi^{a\dagger} \, .
\eea
By complete analogy with our previous discussion of the definition 
of the electric and magnetic charges  
(\ref{q}) and (\ref{g-integral}) above,the volume integrals here can be written 
as the integrals over the surface of the sphere $S^2$ on the spatial infinity:
\begin{equation}
\begin{split}
\int d^3x~F^a_{0k}  \nabla_k \phi^{a\dagger} &= \int d^2 S_k E^a_k \phi^{a\dagger} = 
\int d^3x~ \partial_k(E_k^a \phi^{a\dagger})\, ,\\
\int d^3x~{\widetilde F}^a_{0k}  \nabla_k \phi^{a\dagger} &= \int d^2 S_k B^a_k \phi^{a\dagger} = 
\int d^3x~ \partial_k(B_k^a \phi^{a\dagger}) \, .
\end{split}
\end{equation}
Hence, the anticommutator (\ref{comm-supersurrents}) is of the form 
\begin{equation}
\{Q^1_\alpha,Q^2_\beta\} = - 2 {\sqrt 2}\varepsilon_{\alpha\beta}  
\int d^3x~\partial_k (i E_k^a + B_k^a) \phi^{a\dagger} =  2 \varepsilon_{\alpha\beta} Z \, .
\end{equation}
By the same token, we find that 
\begin{equation}
\{{\bar Q}^1_{\da},{\bar Q}^2_{\db}\} = - 2 {\sqrt 2} \varepsilon_{{\da}{\db}}   
\int d^3x~\partial_k (- i E_k^a + B_k^a) \phi^{a} = 2  \varepsilon_{{\da}{\db}} Z^* \, .
\end{equation}
Here we make use of the definition of the complex 
central charge $Z$ given by Eq~(\ref{z-algebra}). Thus the central charge satisfies 
\begin{equation}    \label{N2centralZ}
Z = {\sqrt  2} \int d^3x~ \partial_k (i E_k^a + B_k^a) \phi^{a\dagger} \, .
\end{equation}
The final step is to recall that the electric and the 
magnetic charges are given by Eq.~(\ref{SUSY-q-g}), where the complex 
scalar field is decomposed into two real components as  
$\displaystyle \phi^a = \frac{1}{\sqrt 2}
(\phi_1^a + i \phi_2^a)$. Therefore, the central charge of the $N=2$ SUSY Yang-Mills 
theory is simply \cite{Witten78} 
\begin{equation}    \label{N2_Z}
\begin{split}
Z &= \int d^3x~ \partial_k ( E_k^a\phi_2^a + B_k^a \phi_1^a) + i  \int d^3x~ \partial_k (E_k^a \phi^a_1 - B_k^a\phi^a_2)\\
&= [(q^1 - g^2) +  i (g^1 + q^2)] \, .
\end{split}
\end{equation}
Thus, the algebra of supersymmetry   (\ref{z-algebra}), which includes the central charge, 
yields a mass bound 
\be \label{BPS-N2}
 |Z| = |(q^1 - g^2) +  i (g^1 + q^2)|   \le M \, ,
\ee
which is precisely the BPS bound (\ref{BPS-bound-11}). Thus, in the $N=2$ SUSY Yang--Mills theory 
the magnetic and the electric charges of
the bosonic monopole configuration appear in the explicit form of the central charge of the  
$N=2$ supersymmetry algebra and the Bogomol'nyi bound is a direct consequence of the extended supersymmetry.  
There is a difference from its classical counterpart (\ref{BPS-bound}), because if the $N=2$ supersymmetry 
is not broken by the one-loop quantum corrections, the Bogomol'nyi bound  (\ref{BPS-N2}) is not 
modified. Note that in that case, the BPS states with magnetic and electric 
charges will be presented in the spectrum of 
physical states of the quantum supersymmetric theory\footnote{It must be kept in mind, 
however, that this statement is correct 
if the vacuum expectation value of the scalar field is large.
Seiberg and Witten pointed out \cite{WS} 
that in the strong coupling limit, the expression (\ref{BPS-bound-11}) must be modified}. 

\subsection{Fermionic Zero Modes in Supersymmetric Theory}
Let us assume that the fields entering the supersymmetry transformations 
(\ref{N=2-Dirac-trans}) satisfy the BPS equations (\ref{BPS-eq-SUSY}). 
Thus, we shall consider the supersymmetry variations on the $SU(2)$ monopole background. 
This is the classical solution of Eq.~(\ref{BPS-Sol}) with no fermions, which we choose 
as an initial configuration. 

To simplify our calculations, we recall that on a classical level, 
the pseudoscalar field 
$\phi^a_2$ can always be eliminated by a chiral rotation. 
Then the supersymmetry 
variations of the bosonic fields in (\ref{N=2-Dirac-trans}) are vanishing and  
the supersymmetry variation of the spinor field becomes 
$$
\delta \chi^a = \biggl(\sigma^{\mu\nu} F^a_{\mu\nu} + \gamma^\mu~ 
(\nabla_\mu \phi_1^a)\biggr)\xi \, . 
$$
 
For a static 
field configuration ($E^a_n = 0$), the BPS equations become simply 
$B_n^a = \nabla_n \phi^a_1$. Then the supersymmetry variation takes the form  
$$
\delta \chi^a = \biggl(\varepsilon_{mnk} \sigma^{mn} B_k^a + 
\gamma^n B_n^a\biggr)\xi = 
 B_n^a \left(\frac{i}{2} \varepsilon_{nmk} \gamma^m \gamma^k + \gamma^n\right)\xi \, , 
$$ 
where we make use of the definition 
\begin{equation}                       \label{sigma-generators}
{(\sigma^{\mu\nu})_\alpha}^\beta = 
\frac{i}{2}\left[(\sigma^\mu)_{\alpha 
{\dot \gamma}} ({\bar \sigma}^\nu)^{{\dot \gamma}\beta} - 
(\sigma^\nu)_{\alpha 
{\dot \gamma}} ({\bar \sigma}^\mu)^{{\dot \gamma}\beta}\right] \, ,
\end{equation}
and related properties of the 
$\gamma$-matrices. Thus, we obtain 
\begin{equation}        \label{susy-bps-var}
\delta \chi^a = \gamma^n B_n^a \left(1 + \frac{i}{3!}\varepsilon_{nmk}\gamma^n \gamma^m 
\gamma^k\right)\xi = \gamma^n B_n^a \left(1 + i\gamma_0\gamma_5\right)\xi \, . 
\end{equation}
 
This means that if the parameter of the supersymmetry transformation $\xi$ satisfies the 
equation 
$$ 
(1 + i\gamma_0\gamma_5)\xi =  (1 - \Gamma_5)\xi = 0 \, ,
$$
where 
$$
\Gamma_5 = -i\gamma_0\gamma_5 = \gamma_1\gamma_2\gamma_3 = 
\begin{pmatrix}0&-i\\i&0
\end{pmatrix} \, ,
$$
and $\Gamma_5^2 = 1$, $\Gamma_5^\dagger = \Gamma_5$,
the variation of the spinor field vanishes identically. 
This corresponds to unbroken
supersymmetry, since we suppose that there are no fermions in the initial configuration. 
Evidently, if we decompose 
$\xi = \xi_+ + \xi_-$, where 
$$
\xi_\pm = \frac{1}{2}\left(1 \pm \Gamma_5\right)\xi\, , 
$$
the transformation of the supersymmetry generated by the parameter $\xi_+$ acts  
on the bosonic monopole background trivially, i.e., $\delta_{\xi_+} \chi^a = 0$.  
On the other hand, the supersymmetry 
variation generated by $\xi_-$  breaks down  half of the
supersymmetry\footnote{This is why this solution sometimes is referred to as 
the {\it 1/2-BPS monopole}.} and drives the configuration from $\chi^a  = 0$ to 
\begin{equation}         \label{susy-ferm-zero}
\delta_{\xi_-} \chi^a \equiv \chi^a_{(0)} =  -2  \gamma^n B_n^a \xi_- \, .
\end{equation}
These zero energy Grassmannian variations of the bosonic monopole solution (\ref{BPS-Sol}) are 
two {\it fermionic zero modes}. 
One can prove that these modes are time-independent solutions 
of the Dirac equation for a fermion coupled with a supersymmetric monopole. 
Indeed, variation of the Lagrangian (\ref{L-N=2-Dirac}) with respect to the 
field ${\bar \chi}$ yields the Dirac equation
$$
i\gamma^\mu \nabla_\mu \chi - e [\phi_1, \chi] = 0 \, .
$$ 
It is straightforward now to substitute 
the explicit form of the fermionic zero modes (\ref{susy-ferm-zero}) into this equation to 
ensure that they are the solutions with zero eigenvalues.  

Let us note that we can expect this effect in advance, 
since the Callias index theorem \cite{Callias78} 
predicts exactly two fermion zero modes for the fermions in the adjoint representation 
of the gauge group. Actually, from the $N=2$ supersymmetry algebra with central 
charge, we already encounter the partial breaking of 
supersymmetry for the states that belong to the short multiplets 
and saturate the BPS bound (see, e.g., \cite{Bilal00,Lykken96}).

The presence of two fermionic zero modes on the monopole bosonic background 
implies that there is {\it N = 2 BPS monopole multiplet}, which can be constructed 
starting from the vacuum spin-0 state $| \Omega \rangle$ by consequent action of the 
operators of creation of these zero modes $a^\dagger_{\pm 1/2}$.  This monopole 
multiplet contains four states: two scalars and two fermions \cite{Osborn-79}.
Note that these 
states are dual to the states of the massive short $N=2$ chiral multiplet of four helicity
states \cite{Bilal00,Lykken96}.
  
Furthermore, the remaining half of the supersymmetry of the $N=2$ BPS monopoles 
allows us to set a correspondence between two fermionic zero modes and four 
bosonic zero modes \cite{Gauntlett94}, which form a supermultiplet with respect 
to the unbroken supersymmetry. Indeed, according to (\ref{N=2-Dirac-trans}),  
for each fermionic zero  mode $\chi_{(0)}$ of Eq.~(\ref{susy-ferm-zero}) 
the remaining half of the supersymmetry transformation  
generated by the supertranslations $\xi_+$ yields  
\begin{equation}    \label{boson-fermion}
\begin{split}
\delta A^a_n &= i{\xi^\dagger}_+ \gamma_n \chi_{(0)}^a - 
i\chi^{a\dagger}_{(0)} \gamma_n \xi_+ \, ,\\
\delta \phi_1^a &= i{\xi^\dagger}_+ \gamma_0 \chi^a_{(0)}- 
i\chi^{a\dagger}_{(0)} \xi_+ \, .
\end{split}
\end{equation}

\subsection{$N=2$ Supersymmetric Monopoles beyond $SU(2)$}
To generalize our previous discussion, let us analyze the properties of 
$N=2$ supersymmetric monopoles in the model with gauge group $G = SU(N)$.

Since we are considering the BPS monopoles, the 
classical potential of the scalar field must vanish. Recall that for $N=2$ SUSY, this
condition is satisfied, if the two components of the Higgs field $\phi_1, \phi_2$ 
are in the Cartan 
subalgebra of $G$ and on the spatial asymptotic
\begin{equation}   \label{phi-H}
\phi_i = ({\vec h}_i \cdot {\vec H}) \, ,  
\end{equation}
where ${\vec h}_i$, $i=1,2$, are vectors in the root space of the Cartan subalgebra 
of dimension $r = ~{\rm rank}~(G)$. 
Then, for a given vacuum, we can also define the electric and magnetic  
charges as vectors in the root space (cf. Eqs. (\ref{wein-quant}) and 
(\ref{SUSY-q-g})) by  
\bea    \label{charges-vect-ch-11}
q_i &=& \int d^3x~ \partial_n \left(E_n^a \phi^a_i\right) 
=  ({\vec q} \cdot {\vec h}_i  )\, ,\nonumber\\ 
g_i  &=& \int d^3x~ \partial_n \left(B_n^a \phi^a_i\right)
= ({\vec g} \cdot {\vec h}_i) \, .
\eea
In this notation, the BPS bound (\ref{BPS-bound-11}) can be written as 
\begin{equation}  \label{BPS-comlex}
M =  \left| ({\vec h}_1 + i {\vec h}_2) \cdot ({\vec q} + i{\vec g})\right|
= {\sqrt 2}~ | {\vec h} \cdot  ({\vec q} + i{\vec g})| \, ,
\end{equation}
where, according to the definition (\ref{phi-H}), the complex vector of the Higgs field is  
$$
{\vec h} = \frac{1}{\sqrt 2}\left( {\vec h}_1 + i {\vec h}_2 \right) \, . 
$$

So far, we have paid little attention to the fact that the electric and magnetic 
vectors lie on two different lattices: the former 
can be expanded in the basis of simple roots ${\vec \beta}_i$ of the given Lie group $G$,  
while the latter is defined in terms of the expansion in the basis of dual co-root vectors  
$\vec\beta_i^*$ of the dual lattice:
\begin{equation}    \label{root-susy-expan}
{\vec q} = e\sum\limits_{i=1}^{r}m_i \vec\beta_i\, ;\qquad 
{\vec g} =\frac{4 \pi}{e}\sum\limits_{i=1}^{r}n_i \vec\beta_i^* \, . 
\end{equation}
Here the integers $m_i,n_i \in \mathbb{Z}$ are the electric\footnote{If we restrict our 
consideration to the classical limit, the electric charge remains non-quantizable.} 
and the magnetic quantum numbers\footnote{For a special unitary group, we can always 
choose the self-dual basis of the simple roots: $\vec\beta_i^* = \vec\beta_i$.}.   

This expansion allows us to write the BPS mass bound in the form  
\bea   \label{BPS-Witten-mod}
M &=& \left| e \sum\limits_{i=1}^{r} m_i \left( {\sqrt 2} \vec\beta_i \cdot {\vec h} 
\right)
+ \frac{4i \pi }{e}\sum\limits_{i=1}^{r} n_i \left( {\sqrt 2} \vec\beta_i^* \cdot {\vec h} 
\right) \right| \nonumber\\
&\equiv& \left| \sum\limits_{i=1}^{r}(m_i \Phi_i + n_i \Phi^*_i) \right| \, ,
\eea
where we introduce the rescaled scalar field and its dual as  
\begin{equation}   \label{aligh-fields}
\begin{split}
\Phi_i = e  {\sqrt 2} \left(\vec\beta_i \cdot {\vec h} \right)\, , \qquad
\Phi^*_i = \frac{4i \pi}{e} {\sqrt 2} \left(\vec\beta_i^* \cdot {\vec h} \right) \, .
\end{split}
\end{equation}
These fields are expanded over the basis of simple roots and co-roots, respectively. 
This form of the 
BPS mass bound is evidently symmetric with respect to the dual transformations. Actually this 
is the Seiberg--Witten form of the BPS boundary, which appears in the quantum $N=2$ 
supersymmetric theory. We shall consider this remarkable relation in more detail 
when we discuss the quantum vacuum moduli space of the $N=2$ SUSY Yang--Mills theory.  

Let us recall now that the BPS equation for the $N=2$ supersymmetric monopoles 
is written in terms of the 
$SO(2)$ rotated Higgs fields ${\tilde \phi}^a_i$ (\ref{rot-higgs}). 
Since the scalar fields lie in the root space of the Cartan subalgebra, we can describe this 
transformation as a rotation of the vectors ${\vec h}_i$, that is,  
$\phi_i \to {\tilde \phi}_i = {\vec h'}_i \cdot {\vec H}$,  where 
\begin{equation} 
{\vec h'}_1 = {\vec h}_1 \sin \delta + {\vec h}_2 \cos \delta,
\qquad  {\vec h'}_2 = {\vec h}_1 \cos \delta - 
{\vec h}_2 \sin \delta \, .
\end{equation}
Asymptotically, these fields  decay  as 
\begin{equation}                \label{asymp-H-decay-ch11}
\begin{split}
{\tilde \phi}_1 (r) &=  
{\vec h'}_1 \cdot {\vec H} - 
\frac{{\vec q} \cdot {\vec H}}{r} + O(r^{-2}),\\
{\tilde \phi}_2 (r) &=  
{\vec h'}_2 \cdot {\vec H} - 
\frac{{\vec g} \cdot {\vec H}}{r} + O(r^{-2}) \, ,
\end{split}
\end{equation}
and the angle of rotation of the Higgs fields is 
restricted by the constraint (\ref{BPS-constr}) 
$$
\tan \delta = \frac{q^1 - g^2 }{g^1 + q^2} \, .
$$
Note that for a purely magnetically charged state, the long-range scalar interaction is 
entirely given by the asymptotic behavior of the field ${\tilde \phi}_2$, while the second component
${\tilde \phi}_1 (r)$ has no Coulomb tail at all. However, in the strong coupling limit, the roles 
of the components of the Higgs field are inverted.  

Taking into account the definitions of the charges (\ref{charges-vect-ch-11}), we can easily see  
that the constraint on the angle of rotation becomes simply 
$$
{\vec g} \cdot {\vec h'}_1 = {\vec q} \cdot {\vec h'}_2 \, ,
$$
and the BPS mass formula (\ref{BPS-delta-rot}) can be written as 
$$
M = \mid {\vec q} \cdot {\vec h'}_1 + {\vec g} \cdot {\vec h'}_2 \mid \, .
$$
These two contributions to the mass are referred to as the 
{\it magnetic mass} $( {\vec g} \cdot 
{\vec h'}_2)$ and the {\it electric mass} $({\vec q} \cdot {\vec h'}_1)$, respectively.
In the weak coupling regime $e \ll 1$, the electric mass is obviously much smaller than
the magnetic mass. This observation justifies the use of the semiclassical 
low-energy approximation.\label{mag-mass}

Let us consider the classical limiting case of vanishing electric mass.
From our previous discussion of the non-supersymmetric $SU(N)$ monopoles, 
we know that the physical situation strongly depends 
on the character of the symmetry breaking. If the vector ${\vec h'}_2$ is not 
orthogonal to any of the simple co-roots  $\vec\beta_i^*$, the vacuum expectation value 
of the scalar field ${\tilde \phi}_2$ breaks the $SU(N)$ symmetry down to 
the residual group $U(1)^{N-1}$ and each of the integers $n_i$ appearing in 
Eq.~(\ref{root-susy-expan}) has the meaning of a topological charge. In this classical 
limit, there are no electric charges of the BPS states and the field  ${\tilde \phi}_1$
does not have a long-range Coulomb tail. 
 
In the particular case of the $N=2$ SUSY Yang-Mills theory with the gauge group 
$SU(3)$, this situation 
corresponds to the $(n_1,n_2)$ 1/2-BPS monopole discussed 
in \cite{Gauntlett00,Lee98}.
In the more general case of the $SU(N)$ gauge group, there are $r=N-1$ types of monopoles and 
the magnetic mass of the corresponding configuration is of the form
$$
M = \left| \frac{4 \pi}{e}\sum\limits_{i=1}^{r}n_i \left( \vec\beta_i^* \cdot {\vec h'}_2 \right)
\right| \, .
$$
Due to triangle inequality, this mass obeys 
$$
M \le \sum\limits_{i=1}^{r}n_i M_i \, ,
$$
where $M_i = {4\pi}(\vec\beta_i^* \cdot {\vec h'}_2)/e $  and, as before, 
we suppose that the set of simple co-roots satisfies the condition  
$ (\vec\beta_i^* \cdot \vec h'_i)  \ge 0$ for all $i$.

Again, we can interpret  $M_i $ as a mass of a single fundamental
monopole with a minimal magnetic charge. 
Indeed, let us recall that 
the magnetic charge satisfies the condition of the topological quantization 
(\ref{top-quantization}):
$\exp\{2i \pi e {\vec g} \cdot {\vec H}\} = 1 $ and the charge matrix is
$$
{\vec g} \cdot {\vec H}  = 
\sum\limits_{i=1}^{r} n_i (\vec\beta_i^* \cdot {\vec H}) =
~{\rm diag}~(k_1, k_2, \dots k_{N-1}) \, ,
$$
with non-negative integers $k_r$, which, in the case of the maximal symmetry breaking, 
are related to the corresponding topological charges.
Hence, the configuration of the mass $M$ is stable with respect to decay into 
$N-1$ species of the fundamental monopoles, each   
of the mass $M_i$ which is associated with the simple co-root  $\vec\beta_i^*$. Even in the  
special case of so-called {\it marginal stability}, when $M =\sum\limits_{i=1}^{r}n_i M_i$, there is no phase space for a 
physical decay. We are already familiar with a similar conclusion in the particular case 
of the composite $SU(3)$ monopole, which is also valid for non-BPS monopoles \cite{S03}. 

So far we have discussing the solutions of the primary BPS equation. 
However, there are solutions of the secondary BPS equation (\ref{second-BPS}), 
which we can find for each solution of the primary
BPS equation. 
 
Let us recall now that the secondary BPS equation is actually the equations for the 
gauge-orthogonal zero modes,  
which corresponds to the  
large gauge transformations of the fields of the BPS monopole. 
The latter solution of the primary BPS equation corresponds to an 
electrically neutral configuration. If the gauge symmetry is broken
maximally, there are $N-1$ such gauge zero modes. Solving the secondary BPS equation 
(\ref{second-BPS}), 
we can recover the electric charges of the monopoles 
from the asymptotic of the Higgs fields (\ref{asymp-H-decay-ch11}), 
in other words, we are ``dressing the monopole electrically''.  

However, the vacuum expectation value of the ``electric'' component of the Higgs field ${\tilde \phi}_1$
is no longer obliged to be proportional to the  ``magnetic'' component ${\tilde \phi}_2$, 
as happens in the case 
of the $SU(2)$ supersymmetric Yang--Mills theory. In other words, the  electric charge 
vector ${\vec q}$ 
of an $SU(N)$ dyon is no longer aligned with the magnetic charge vector ${\vec g}$.
 
Typically, in theories with extended $N=4$ supersymmetry, the dyonic BPS states, which are 
solutions both of the primary and the secondary BPS equations, 
break 3/4 of the supersymmetry\footnote{Recall that field content of $N=4$ supersymmetric
theory includes 
six real scalar fields. Thus, we have to set four of these fields to be zero to 
reduce the Lagrangian to the form (\ref{L-N=2-Dirac})}, while in the model with 
$N=2$ supersymmetry, they still preserve
half of the supersymmetry \cite{Bak99,Bak00-1,Bak00-2,Gauntlett01,Lee98,Ritz00,Tong99}.  
Nevertheless, somewhat inconsistently, 
they are referred to as the {\it 1/4-BPS states}. The interpretation of the 
solutions of $SU(N)$ BPS equations as a composite system of $N-1$ fundamental monopoles,  
suggests that such a 1/4-BPS configuration can be thought of as a static system of a few 
1/2-BPS monopoles. These solutions correspond to the composite root vectors of the 
Cartan--Weyl basis. 

Indeed, a fundamental magnetic monopole, which corresponds to a 
simple co-root ${\vec \beta}^*_i$, is a solution of the primary BPS equation. Each such a 
monopole could have only its own type of electric charge, which corresponds to the 
root ${\vec \beta}_i$. The self-duality of the basis means that for a fundamental $SU(N)$ 
monopole, the electric and the magnetic charge vectors are aligned. 

The situation is different in the case of the composite monopoles,    
which correspond to the composite roots: they consist of two or more
fundamental monopoles on top of each other. Recall that this configuration is static,  
because the electric
(Coulomb) part of the interaction between the monopoles is precisely compensated for 
by the long-range scalar force. Then the electric charges of the different monopoles 
are functions of their relative orientation. 
Hence the low-energy dynamics
of these BPS states becomes more complicated than in the case of the simple non-supersymmetric $SU(2)$ 
gauge theory, because the low-energy Lagrangian of the composite monopoles 
picks up an additional term, which is associated with different orientations of two 
scalar fields. We shall consider this situation below.

To recover the low-energy effective Lagrangian of the supersymmetric monopoles 
we can make use of the same approach as in our consideration above.It was argued \cite{Bak00-1,Lee98,Tong99} that
the corresponding potential is simply half of the electric mass of the configuration, 
that  is,  
$$
V_{eff} = \frac{1}{2} \left({\vec q} \cdot {\vec h}'_1\right) \, .
$$
Recall that  in the weak coupling regime this is a small correction to the magnetic mass. 
Taking into account the definition of the electric charge vector (\ref{charges-vect-ch-11})
and the equation of motion of the field ${\tilde \phi}^a_1$ (the secondary BPS equation 
(\ref{second-BPS})), 
we can write the electric mass as \cite{Tong99}
\begin{equation}
\begin{split}
\left({\vec q} \cdot {\vec h}'_1\right) &= \int d^3x~ \partial_n \left(E_n^a 
{\tilde \phi}^a_1\right) = \int d^3x~ \partial_n  \left( {\tilde \phi}^a_1 \nabla_n 
{\tilde \phi}^a_1 \right)\\
&=  \Tr \int d^3x~ \biggl\{ (\nabla_n{\tilde \phi}_1 )^2 - e^2 [ {\tilde \phi}_1,  
{\tilde \phi}_2] \biggr\} \, .
\end{split}
\end{equation}
However, $(\nabla_n{\tilde \phi}_1 )$ is a large gauge transformation of the monopole field 
with the gauge parameter ${\tilde \phi}_1$. These transformations 
correspond to the set of gauge zero modes of the configuration:
\begin{equation}
\begin{split}
\delta A_n &= \nabla_n {\tilde \phi}_1 = \sum\limits_\alpha \left( {\vec h}'_1 
\cdot {\vec K}^\alpha\right) \delta_\alpha A_n \equiv G^\alpha 
\delta_\alpha A_n\, ,\\
\delta {\tilde \phi}_2 &= ie [{\tilde \phi}_1, {\tilde \phi}_2] = 
\sum\limits_\alpha \left( {\vec h}'_1 
\cdot {\vec K}^\alpha \right) \delta_\alpha {\tilde \phi}_2 \equiv 
G^\alpha \delta_\alpha {\tilde \phi}_2 \, , 
\end{split}
\end{equation}
where $ {\vec K}^\alpha$ are the components of the Killing vector 
field $G = ( {\vec h}'_1 
\cdot {\vec K}^\alpha )$ on the moduli space 
${\cal M}$, which are generated by the $U(1)^r$ gauge transformations. Thus, the electric 
mass of the 1/4-BPS state can be written as 
\begin{equation}
\left({\vec q} \cdot {\vec h}'_1\right) = g_{\alpha \beta} \left({\vec q} \cdot 
{\vec K}^\alpha\right) \left({\vec q} \cdot  
{\vec K}^\beta\right) \, ,
\end{equation}
where $ g_{\alpha \beta}$ is the hyper-K\"ahler metric on ${\cal M}$. 
If the form of this metric is known, as in the particular case of  
the $SU(3)$ 1/4-BPS monopole, the electric mass can be calculated directly 
from the  metric. An alternative approach is to write the metric in terms 
of the Nahm data \cite{Houghton00}.

\subsection{$SU(3)$ $N=2$ Supersymmetric Monopoles}
\begin{figure}   
\centering
\includegraphics[height=8.cm]{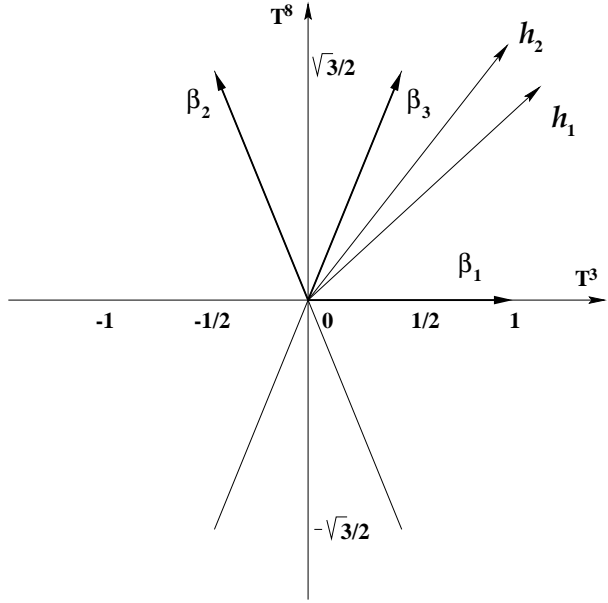}
\caption{
Positive simple roots of the $SU(3)$ supersymmetric 
theory. The vectors ${\vec h}_i$ define the 
orientation of two scalar fields. 
\label{fig11.1} }
\end{figure}

We can proceed further by analogy with our previous discussion, 
where we concentrated on the particular case of the $SU(3)$ gauge theory. 
Then the corresponding Cartan subalgebra is given by two generators  (\ref{H-SU3-ch8})
and the self-dual basis of the simple 
roots can be chosen in the form (\ref{simple-root}) as before: 
\begin{equation}           \label{simple-root-ch11}
{\vec \beta}_1 = (1, 0) ,\qquad {\vec \beta}_2 = (-1/2, {\sqrt 3}/2) \, .
\end{equation}
In addition, there is the third, composite root, which is defined 
as $ \vec \beta_3 = \vec \beta_1 + \vec \beta_2 $, as shown in 
Fig.~\ref{fig11.1}.
Thus, $\vec \beta^*_i = \vec \beta_i$. 

Let us consider the case of the maximal symmetry breaking: 
$SU(3) \to H = U(1)\times U(1)$. 
Then there are are two fundamental monopoles with vector magnetic charges 
${\vec g} = (1,0)$ and 
${\vec g} = (0,1)$, which are aligned along fundamental roots 
${\vec \beta}_1$ and ${\vec \beta}_2$, respectively. The corresponding charge matrices 
${\vec g} \cdot {\vec H}$ are 
$$
\frac{1}{2}~{\rm diag}~(1,-1,0),\qquad \frac{1}{2}~{\rm diag}~(0,1,-1) \, ,
$$
while the charge matrix of the composite  $(1,1)$ monopole is 
$\frac{1}{2}~{\rm diag}~(1,0,-1)$. The latter state is 
the 1/4-BPS configuration that we would like to consider.

Let us take the electric  
numbers of these states to be $m_i = (q_1/e, q_2/e)$, where $q_i$ are arbitrary numbers.
The magnetic charge of the fundamental monopole is $g = 4\pi /e$ and, 
according to Eq.~(\ref {BPS-Witten-mod}), these states have the masses
\begin{equation}   \label{two-BPS-masses}
M_i = {\sqrt 2} |(q_i + ig) ( \vec\beta_i \cdot {\vec h})| =  
{\sqrt 2} \left|\Phi_i\right|\sqrt{q_i^2 + g^2} \, , 
\end{equation}
which evidently corresponds to Eq.~(\ref{mass-su}). Recall that 
in the weak coupling limit, the masses  (\ref{two-BPS-masses}) are slightly different
from the mass of a fundamental electrically neutral BPS-state, which is now given by  
${\sqrt 2}g \left|\Phi_i\right|$.

The complex scalar fields $\Phi_1$ and $\Phi_2$ defined as in Eq.~(\ref{aligh-fields}) 
are now disaligned, that is,  
$$
\Phi_i = e {\sqrt 2} \left|  ( \vec\beta_i \cdot {\vec h}) \right| e^{i \omega_i} \, , 
$$
where 
$$
\left| \Phi_i\right| =  
e  {\sqrt 2} \left|  ( \vec\beta_i \cdot {\vec h}) \right| = 
e \sqrt{ ({\vec \beta_i} \cdot {\vec h}_1)^2 +  ({\vec \beta_i} \cdot 
{\vec h}_2)^2} \, ,
$$
and the agrument of the complex Higgs field $\Phi_i$ is 
$$
\tan \omega_i = \frac{({\vec \beta_i} \cdot 
{\vec h}_2)}{({\vec \beta_i} \cdot 
{\vec h}_1) } \, .
$$

In addition, there is a scalar  charge of the Higgs field, which is defined according 
to Eq.~(\ref{dilat-su}) as
\begin{equation}        
Q_D^{(i)} = ({\vec \beta}_i \cdot {\vec H}) \sqrt{q_i^2 + g^2} \, .
\end{equation}  
Hence, the scalar part of the long-range interaction between 
the monopoles depends on the relative orientation of the Higgs fields: 
it vanishes if the 
fields $\Phi_1$ and $\Phi_2$ are anti-parallel, while its magnitude becomes maximal if 
they are aligned. 
Thus, by analogy with Eq.~(\ref{misalign}),  
we can now write the total Coulomb potential  of the composite $(1,1)$ monopole,
which consists of two static components \cite{Bak00-1,Ritz00}:
\begin{equation}   \label{pot-dil-ch-11}
V_{eff} =   - \frac{1}{r}\left(q_1q_2 + g^2 -  \cos(\omega_1 - \omega_2) 
\sqrt{(q_1^2 + g^2)(q_2^2 + g^2)}\right) \, .
\end{equation}
Here we also take into account electrostatic and magnetostatic contributions. 

Note that the mass of the composite ${\vec g} = (1,1)$ 1/4-BPS state  
\be       \label{mass4}
M_{(1,1)} = {\sqrt 2}| (q_1 + ig) ( \vec\beta_1 \cdot {\vec h})
+ (q_2 + ig) ( \vec\beta_2 \cdot {\vec h})| 
\ee
becomes additive, i.e., $M_{(1,1)} = M_1 + M_2$, only if \cite{Ritz00} 
\begin{equation}    \label{disalign}
\tan (\omega_1 - \omega_2) = \frac{g(q_1 - q_2)}{q_1q_2 + g^2} \, . 
\end{equation}
In this case, the dilatonic part of the Coulomb interaction in Eq.~(\ref{pot-dil-ch-11})
precisely balances the long-range electromagnetic interaction and the potential vanishes.
In other words, two constituent monopoles are static and the 1/4-BPS configuration can be 
regarded as a superposition of two individual 1/2-BPS states. 
  
If the difference $\Delta\omega = \omega_1 - \omega_2$ 
of the arguments of the complex fields 
$\Phi_i$ is small, we can easily see that the expansion of 
the Coulomb potential (\ref{pot-dil-ch-11}) 
in $q_i/g$ yields
$$
V_{eff} \approx \frac{1}{2r} \left[Q^2 - g^2 (\omega_1 - \omega_2)^2\right] \, ,
$$
where a relative electric charge of the BPS states is $Q = q_1 - q_2$.
Clearly, this potential is repulsive if $| g\Delta\omega /Q| < 1$, that is, 
in this case the supersymmetric 1/4-BPS monopole does not exist. The net interaction 
is vanishing if   
$$
\Delta\omega = \omega_1 - \omega_2 = \frac{Q}{g} \, .
$$
This is a condition of stability of the 1/4-BPS monopole, which evidently 
agrees with Eq.~(\ref{disalign}). 
  
Let us briefly describe the low-energy dynamics of the 1/4-BPS monopoles.  
We have seen that, for an arbitrary orientation of the component of the Higgs field, there is a 
non-vanishing potential of the interaction $V_{eff}$. This potential may be 
attractive and a 1/4-BPS monopole exists as a bounded system of two fundamental monopoles.  
Thus, the low-energy approximation can be applied only if the potential energy is small compared
to the rest mass of the fundamental monopole.

To sum up, the long-distance tail 
of the Higgs field ${\tilde \phi}_1$, which is associated with the electric charge vector
${\vec q}$, gives rise to the potential of interaction $V_{eff}$.   
Then the low-energy 
effective Lagrangian which describe the relative bosonic collective coordinates 
$X^\alpha$ of the 1/4-BPS state, 
besides the usual kinetic term includes the potential piece 
\cite{Bak00-1,Tong99,Gauntlett00,Gauntlett01}.

So far, we have been concerned about maximal symmetry breaking. 
The case of minimal $SU(3)$ symmetry breaking was considered 
recently in the paper \cite{Houghton00}. Recall that then the Higgs 
field ${\vec h}'_2$ is orthogonal to 
one of the simple roots, say ${\vec \beta_1}$, and its asymptotic value
breaks the symmetry to $U(2)$.  The corresponding spectrum of states   
includes the massive fundamental ${\vec \beta_2}$-monopole with a topological 
charge $n$  
and the massless ${\vec \beta_1}$-monopole.  

The configuration of the $(2,[1])$ supersymmetric monopole 
was considered in \cite{Houghton00}. This is an example of 
two massive monopoles and one massless monopole, which can be thought of as a non-Abelian cloud 
surrounding the two massive monopoles. 
The difference from the non-supersymmetric case is that 
there is now a long-range tail of the 
second scalar field ${\vec h}'_1$, which  
breaks the symmetry further to the minimal subgroup $U(1) \times U(1)$. 
It was shown that the potential of the $(2,[1])$ supersymmetric monopole 
is attractive and the massless monopole is confined to one of these two massive 
monopoles. 

Let us stop our discussion at this point. 
Recent developments in the understanding of the low-energy dynamics of the 
supersymmetric monopoles, which basically used the same simple picture 
of geodesic motion on the underlying moduli space suggested by N.~Manton
in 1982 \cite{Manton82}, have greatly improved our understanding of the 
structure of the vacuum of supersymmetric theories. The restricted volume  
of our review does not allow us to go into detail of many remarkable 
works. In particular, the general description of the low-energy dynamics 
of the supersymmetric monopoles was given recently in \cite{Gauntlett01},  
where the complete effective Lagrangian of  
bosonic and fermionic collective coordinates was derived. We also do not 
discuss here the powerful Nahm formalism, which allows us to obtain many results in a very 
simple and elegant way \cite{Houghton00,Lee98}. 
In this rapidly developing situation, we 
direct the reader to the original works 
\cite{Bak99-1,Bak00-1,Bak00-2,Bergman98,Bergman98-1,Gauntlett00,Gauntlett01,Houghton00,Lee98,Ritz00,Tong99}.

\paragraph{Acknowledgements}
The author thanks the organizers of the 
Advanced Summer School on Modern Mathematical Physics (JINR Dubna, July 2005)
and at the  8th International School-Seminar 
``The actual problems of microworld physics 2005'' (Gomel-Dubna, August 2005),
where a slightly modified version of the lectures was presented, for their invitation and 
kind hospitality.

\end{document}